\magnification 1060
\baselineskip 14 pt

\def \P {{\bf P}}
\def \E {{\bf E}}
\def \F {{\bf F}}
\def \D {{\bf D}}
\def \C {{\cal C}}
\def \L {{\cal L}}
\def \R {{\bf R}}

\def \la {\lambda}
\def \al {\alpha}
\def \be {\beta}
\def \sig {\sigma}
\def \td {\tau_d}
\def \tauo {\tau_0}
\def \noi {\noindent}

\def \phio {\phi^0}
\def \pio {\pi^0}
\def \po {p_0}
\def \phid {\phi^d}

\def \ws {w_s}
\def \wb {w_b}

\def \yo {y_0}
\def \yb {y_b}
\def \yg {y_g}
\def \ygo {y_{g0}}
\def \ybg {y_{bg}}
\def \ybo {y_{b0}}

\def \tw {\tilde w}

\def \tpi {\tilde \pi^*}

\def \sect#1{\bigskip  \noindent{\bf #1} \medskip}
\def \subsect#1{\bigskip \noindent{\it #1} \medskip}

\def \thm#1#2{\medskip \noindent {\bf Theorem #1.}   \it #2 \rm \medskip}
\def \prop#1#2{\medskip \noindent {\bf Proposition #1.}   \it #2 \rm \medskip}
\def \cor#1#2{\medskip \noindent {\bf Corollary #1.}   \it #2 \rm \medskip}
\def \pf {\noindent  {\it Proof}.\quad}
\def \lem#1#2{\medskip \noindent {\bf Lemma #1.}   \it #2 \rm \medskip}
\def \ex#1{\medskip \noindent {\bf Example #1.}}

\def \rem#1{\medskip \noindent {\bf Remark #1.}}

\def \sqr#1#2{{\vcenter{\vbox{\hrule height.#2pt\hbox{\vrule width.#2pt height#1pt \kern#1pt\vrule width.#2pt}\hrule height.#2pt}}}}

\def \square{\hfill\mathchoice\sqr56\sqr56\sqr{4.1}5\sqr{3.5}5}

\def \qed {$\square$ \medskip}

\nopagenumbers

\headline={\ifnum\pageno=1 \hfill \else \hfill {\rm \folio} \fi}

\centerline{\bf Purchasing Term Life Insurance to Reach a Bequest Goal while Consuming}

\bigskip

\centerline{Erhan Bayraktar}
\centerline{Department of Mathematics, University of Michigan}
\centerline{Ann Arbor, Michigan, USA, 48109} \bigskip

\centerline{S. David Promislow}
\centerline{Department of Mathematics, York University}
\centerline{Toronto, Ontario, Canada, M3J 1P3} \bigskip

\centerline{Virginia R. Young}
\centerline{Department of Mathematics, University of Michigan}
\centerline{Ann Arbor, Michigan, USA, 48109} \bigskip

\centerline{Version:  26 February 2016} \bigskip

\noindent{\bf Abstract:}  We determine the optimal strategies for purchasing term life insurance and for investing in a risky financial market in order to maximize the probability of reaching a bequest goal while consuming from an investment account.   We extend Bayraktar and Young (2015) by allowing the individual to purchase term life insurance to reach her bequest goal.  The premium rate for life insurance, $h$, serves as a parameter to connect two seemingly unrelated problems.  As the premium rate approaches $0$, covering the bequest goal becomes costless, so the individual simply wants to avoid ruin that might result from her consumption.  Thus, as $h$ approaches $0$, the problem in this paper becomes equivalent to minimizing the probability of lifetime ruin, which is solved in Young (2004).  On the other hand, as the premium rate becomes arbitrarily large, the individual will not buy life insurance to reach her bequest goal.  Thus, as $h$ approaches infinity, the problem in this paper becomes equivalent to maximizing the probability of reaching the bequest goal when life insurance is not available in the market, which is solved in Bayraktar and Young (2015).

\bigskip


\sect{1. Introduction}

We determine the optimal strategies for purchasing term life insurance and for investing in a risky financial market in order to maximize the probability of reaching a bequest goal while consuming from an investment account.  We extend Bayraktar and Young (2015) by allowing the individual to purchase term life insurance to reach her bequest goal.  The premium rate for life insurance, $h$, serves as a parameter to connect two seemingly unrelated problems.  As the premium rate approaches $0$, covering the bequest goal becomes costless, so the individual simply wants to avoid ruin that might result from her consumption.  Thus, as $h$ approaches $0$, the problem in this paper becomes equivalent to minimizing the probability of lifetime ruin, which is solved in Young (2004).   On the other hand, as the premium rate becomes arbitrarily large, the individual will not buy life insurance to reach her bequest goal.  Thus, as $h$ approaches infinity, the problem in this paper becomes equivalent to maximizing the probability of reaching the bequest goal when life insurance is not available in the market, which is solved in Bayraktar and Young (2015).

The work in this paper combines two areas of research.  One area is the optimal purchase of life insurance.  There are two traditional reasons for buying life insurance: (1) protecting  household income when a wage earner dies, and (2) individuals wishing to leave bequests to their children or other heirs.  To address the first problem, researchers generally assume households wish to maximize utility of consumption and, perhaps, bequest.  For example, Bayraktar and Young (2013) maximize expected exponential utility of a household's consumption when an income earner might die; they determine the optimal investment and life insurance purchasing strategies.\footnote{$^1$}{In a paper with a similar model, but without life insurance available to protect income, Vellekoop and Davis (2009) maximize utility of consumption when income might cease due to a random event.}  Pliska and Ye (2007) have no risky asset in their financial market, and the optimal strategy for purchasing insurance decreases with increasing wealth.  By contrast, Richard (1975) includes a risky asset in his financial market, and the optimal strategy for purchasing insurance increases with wealth.\footnote{$^2$}{In both of those papers, the authors maximize expected utility of consumption plus expected utility of wealth at the time of death.} Thus, the presence of a risky asset can affect the optimal strategy for purchasing insurance; we note a similar difference in Section 3 of this paper.

The other area is the optimal control of wealth to reach a goal.  Research on this topic began with the seminal work of Dubins and Savage (1965, 1976) and continued with the work of Pestien and Sudderth (1985), Orey et al.\ (1987), Sudderth and Weerasinghe (1989), Kulldorff (1993), Karatzas (1997), and Browne (1997, 1999a, and 1999b).  A typical problem considered in this research is to control a process to maximize the probability the process reaches $b$, either before a fixed time $T$, such as in Karatzas (1997), or before the process reaches $a < b$, such as in Pestien and Sudderth (1985).  In either of these forms of the problem, the game ends if wealth reaches $b$.  The problem we consider in this paper is similar in that we control a wealth process to maximize the probability of reaching $b$ before $0$, but we want to reach $b$ at a random time, namely, the time of death of the investor.  The game does not end if wealth reaches $b$ before the investor dies; the game only ends when the individual dies or ruins.

Related goal-seeking research in insurance economics focuses on minimizing the probability of financial ruin of an infinitely-lived insurance company by controlling, for example, investment in a risky financial market and purchasing reinsurance; see Schmidli (2002) and Promislow and Young (2005) for two relatively early papers in this area.  By contrast with that line of research, we focus on individual decision making, which includes the possibility of bankruptcy, as in Schmidli (2002), but which also includes the possibility of death and incorporates life insurance in the market to ``compensate'' the heirs when that event occurs.  If an insurance company were to maximize the probability of being prepared for a catastrophe, then that problem would be more closely related to what we are considering in this paper.

Bayraktar, Promislow, and Young (2014, 2015) introduce the problem of allowing the individual to buy life insurance in order to reach a bequest goal.  In that work, the individual does not consume from the investment account, nor is there a risky asset in which the individual could invest.  The only uncertainty is the time of death.  Bayraktar, Promislow, and Young (2014) consider a time-homogeneous model in which the hazard rate $\la$ and the riskless return $r$ are constant.  We discuss the results of that paper at the beginning of Section 3.  Bayraktar, Promislow, and Young (2015) extend their 2014 paper by allowing the hazard rate to vary deterministically with time, while keeping the remainder of the model the same.  One interesting result from the latter paper is that if the  future lifetime random variable is uniformly distributed on $[0, T]$, then the optimal strategy for purchasing life insurance is as follows: (1) If $r \le {1 \over T}$, then it is optimal for the individual to buy full life insurance until death or ruin, whichever comes first.  (2) If $r > {1 \over T}$, then it is optimal not to buy life insurance until time $T - {1 \over r}$, after which time, it is optimal to buy full life insurance until death or ruin.

Then, Bayraktar and Young (2015) consider the problem of reaching the bequest goal with no life insurance available in the market but, unlike Bayraktar, Promislow, and Young (2014, 2015), the individual consumes from her investment account and there is a risky asset in which the individual could invest.  The main contributions of Bayraktar and Young (2015) are (1) to prove the duality between the bequest problem and an optimal stopping problem in a particular case (see Sections 4.1 and 4.2); and (2) to emend some details of Browne (1997) (see Theorem 4.3 and Remark 4.1).  Also, that paper provides some mathematical background for the methods used in this paper.

In this paper, we extend the work of Bayraktar and Young (2015) by including life insurance in the market.  (Also, we extend Bayraktar, Promislow, and Young (2014) by including a risky asset and non-zero consumption.)  The resulting optimal strategy for purchasing life insurance is {\it very} different from the one found in Bayraktar, Promislow, and Young (2014).  Specifically, when there is no risky asset, if the mortality rate is less than the riskless return, the optimal strategy is not to buy life insurance until wealth reaches the safe level; otherwise, if the mortality rate is greater than the riskless return, the optimal strategy is to buy life insurance if wealth is less than some level.  When there is a risky asset and when the rate of consumption is large enough, it is optimal for the individual to buy insurance at all levels of wealth, a result we did not expect; otherwise, for lower rates of consumption, it is optimal to buy insurance only when wealth is greater than some level.

The rest of the paper is organized as follows. In Section 2, we present the financial and insurance market in which the individual invests, we formalize the problem of maximizing the probability of reaching a bequest goal, and we give a verification lemma that will help us to find that maximum probability, along with the optimal strategies for investing in the financial market and for purchasing term life insurance.  In Section 3, we solve the problem of maximizing the probability of reaching a bequest goal when the rate of consumption is zero; we separate this case because we can solve it explicitly.   Sections 4 and 5 parallel Section 3 for a positive rate of consumption.  In that case, we have an explicit solution only when the rate of consumption is large enough but not too large (Section 4.1). Otherwise, we solve the problem by solving for the convex Legendre transform of the maximum probability and, then, use the verification lemma to prove that our ansatz is, indeed, the maximum probability of reaching the bequest goal.

In Section 6, we prove properties of the solution obtained in Sections 3 through 5.  In particular, we show that, as the premium rate for life insurance $h$ increases, the maximum probability of reaching the bequest goal (weakly) decreases, and the optimal amount invested in the risky asset (weakly) increases.  Furthermore, we show that as $h$ approaches $0$, the maximum probability of reaching the bequest goal and the optimal amount invested in the risky asset approach $1$ minus the minimum probability of lifetime ruin and the corresponding optimal strategy, respectively.  Also, we show that as $h$ approaches $\infty$, the solution to the problem in this paper approaches the solution in Bayraktar and Young (2015) for the problem of maximizing the probability of reaching a bequest goal without life insurance available in the market. Finally, at the end of Section 6, we provide numerical examples to illustrate our results.  Section 7 concludes the paper.

\sect{2.  Statement of the problem and verification lemma}

In this section, we define the financial and insurance market in which the individual invests.  Then, we state the optimization problem this individual faces and present a verification lemma we use in Sections 3 through 5 to solve the optimization problem.

\subsect{2.1. Financial market and probability of reaching the bequest goal}

We assume the individual has an investment account she manages in order to reach a given bequest goal $b$.  She consumes from this account at the constant rate $c$.  The individual invests in a Black-Scholes financial market with one riskless asset earning interest at the rate $r \ge 0$ and one risky asset whose price process $S = \{ S_t \}_{t \ge 0}$ follows geometric Brownian motion:
$$
dS_t = \mu \, S_t \, dt + \sigma \, S_t \, dB_t,
$$
in which $B = \{ B_t \}_{t \ge 0}$ is a standard Brownian motion on a filtered probability space $(\Omega, {\cal F}, \F = \{ {\cal F}_t \}_{t \ge 0}, \P)$, with $\mu > r$ and $\sigma > 0$.

Let $W_t$ denote the wealth of the individual's investment account at time $t \ge 0$.  Let $\pi_t$ denote the dollar amount invested in the risky asset at time $t \ge 0$.  An investment policy $\Pi = \{ \pi_t \}_{t \ge 0}$ is {\it admissible} if it is an $\F$-progressively measurable process satisfying $\int_0^t \pi^2_s \, ds < \infty$ almost surely, for all $t \ge 0$.

Denote the future lifetime random variable of the individual by $\td$.  We assume $\td$ follows an exponential distribution with mean $1/\la$.  We assume the individual buys life insurance via a premium paid continuously at the rate of $h > 0$ per dollar of insurance.   Furthermore, we assume the individual can change the amount of her insurance coverage at any time, that is, the individual may purchase so-called {\it instantaneous} term life insurance.  Bayraktar et al.\ (2014, Section 3.1) assume $h \ge \la$, but we allow $h < \la$ in this paper to account for imperfect information of the insurer in pricing life insurance for a particular individual.

Let $D_t$ denote the amount of death benefit payable at time $\td$ in force at time $t \ge 0$.  An insurance strategy $\D = \{ D_t \}_{t \ge 0}$ is {\it admissible} if it is $\F$-progressively measurable and non-negative.  Thus, with instantaneous term life insurance, wealth follows the dynamics
$$
\left\{
\eqalign{
dW_t &= (r W_t + (\mu - r) \pi_t - c - h \, D_t) \, dt + \sigma \, \pi_t \, dB_t, \quad 0 \le t < \td, \cr
W_{\td} &= W_{\td-} + D_{\td-}.
}
\right.
\eqno(2.1)
$$

We assume the individual seeks to maximize the probability that $W_{\td} \ge b$, by optimizing over admissible controls $(\Pi, \D)$.  Because of the constant drain on wealth by the negative drift term $- c$, financial ruin might occur before death, and we end the game if wealth reaches 0 before the individual dies.  Define $\tauo = \inf \{ t \ge 0: W_t \le 0 \}$, and define the value function by
$$
\phi(w) = \sup_{(\Pi, \D)} \P^{w} \left( W_{\td \wedge \tauo} \ge b \right),
\eqno(2.2)
$$
in which $\P^{w}$ denotes conditional probability given $W_0 = w \ge 0$. We refer to $\phi$ as the maximum probability of reaching the bequest goal, with the understanding that if ruin occurs before death, then the bequest goal cannot be attained.

\rem{2.1} {We assume that the individual's consumption rate $c$ is constant and exogenously given.  If the individual were allowed to control her consumption process $\{ c_t \}$, then, to maximize the probability of reaching her bequest goal, she would optimally choose $c_t \equiv 0$ almost surely, for all $t \ge 0$.  In that case, we argue that the individual would starve to death.  Therefore, we assume that there is a {\it subsistence level}, as in Sethi et al.\ (1992), below which the individual will not reduce her consumption.  The consumption rate $c$ in this paper represents that subsistence level.

The results of this paper will give the optimal investment and life insurance purchasing strategies for a given value of $c$, along with the maximum probability of reaching a bequest goal.  The individual can, then, vary $c$ and compare the resulting strategies and value function to better understand her options.  A natural extension, then, is to allow a time-varying rate of consumption (varying as a deterministic function of time or of wealth), and we invite the interested reader to pursue this problem.  \qed}

%{Because we assume that the parameters of the model are constant, $\phi$ depends only on the state variable $w$. We do not feel that the time independence of the parameters is a drawback of our model.  Indeed, within our relatively simple market, we obtain (semi-)explicit expressions for the optimal strategies for investing and for purchasing life insurance insurance.  Working with simple models helps researchers to determine what properties might be true more generally.  Under more realistic models, we fully expect that our qualitative results will still hold, and perhaps the simple strategies we found will be nearly optimal.  For example, Moore and Young (2006) observed that when minimizing the probability that an individual financially ruins before dying, investment strategies computed by assuming that hazard rates are constant are nearly optimal for the case of time-varying hazard rates.  Bayraktar et al.\ (2011) obtained similar results in a setting of stochastic volatility. \qed}

\rem{2.2} {As the premium rate for life insurance approaches $0+$, then the bequest goal can be covered without cost, and the problem reduces to the one of minimizing the probability of lifetime ruin (Young, 2004).  On the other hand, as the premium rate becomes arbitrarily large, then we expect the individual to buy no life insurance, and the problem reduces to the one of maximizing the probability of reaching the bequest without life insurance in the market (Bayraktar and Young, 2015).  Thus, the problem solved in this paper connects these two seemingly unrelated problems as the premium rate varies from $0$ to $\infty$; see the work in Section 6 below for results proving the continuity of our solution as $h \to 0+$ and as $h \to \infty$. \qed}

\rem{2.3} {If wealth is large enough, say at least $\ws$ (``s'' for safe), then the individual can invest all her wealth in the riskless asset with the interest income sufficient to cover her consumption and insurance premium for death benefit $(b - \ws)_+$.  That is, wealth $\ws$ generates interest of $r \ws = c + h(b - \ws)_+$.  By solving this equation for $\ws$, we obtain
$$
\ws = 
\cases{
{c + hb \over r + h}, &if $0 \le c \le rb$, \cr \cr
{c \over r}, &if $c > rb$,}
\eqno(2.3)
$$
which we call the {\it safe level}.  Thus, $\phi(w) = 1$ if $w \ge w_s$, and it remains for us to determine $\phi(w)$ for $0 < w < w_s$.}

\subsect{2.2 Verification lemma}

In this section, we provide a verification lemma that states that a smooth solution to a boundary-value problem (BVP) associated with the maximization problem in (2.2) equals the value function $\phi$.  Therefore, we can reduce our problem to one of solving a BVP.  We state the verification lemma without proof because its proof is similar to others in the literature; see, for example, Schmidli (2002, Theorem 1), Promislow and Young (2005, Theorem 2.1), or Wang and Young (2012, Theorem 3.1 and Corollary 3.1).

First, for $\pi \in \R$ and $D \ge 0$, define a differential operator $\L^{\pi, D}$ by its action on a test function $f$, whose definition is derived from the wealth dynamics in (2.1):
$$
\eqalign{
\L^{\pi, D} \, f &= (rw + (\mu - r) \pi - c - h \, D) f_w + {1 \over 2} \sigma^2 \pi^2 f_{ww} - \la \left(f - {\bf 1}_{\{ w + D \ge b \}} \right).
}
\eqno(2.4)
$$

\lem{2.1} {Let $\Phi = \Phi(w)$ be a $\C^2([0, \ws])$ function for which $\Phi_w > 0$ and $\Phi_{ww} < 0$ in $(0, \ws)$.  Suppose $\Phi$ satisfies the following BVP on $[0, \ws]:$
$$
\left\{
\eqalign{
&\max_{\pi, D \ge 0} \L^{\pi, D} \, \Phi(w) = 0, \cr
&\Phi(0) = 0, \quad \Phi(\ws) = 1.
}
\right.
\eqno(2.5)
$$
Then, on $[0, \ws]$,
$$
\phi = \Phi,
$$
the optimal amount invested in the risky asset is given in feedback form by
$$
\pi^*_t = - {\mu - r \over \sig^2} \, {\Phi_w(W^*_t) \over \Phi_{ww}(W^*_t)},
\eqno(2.6)
$$
in which $W^*_t$ is the optimally controlled wealth at time $t$, and the optimal amount of instantaneous term life insurance equals
$$
D^*_t = (b - W^*_t) \, {\bf 1}_{\{ \wb \le W^*_t \le \min(\ws, b) \}},
\eqno(2.7)
$$
in which
$$
\wb = \inf \{w \ge 0: \la - h(b - w) \, \Phi_w(w) \ge 0 \} \wedge b.
\eqno(2.8)
$$}

\rem{2.4} {Note that the optimal amount of life insurance to buy at any instant is either 0 or $b - w$, as written in (2.7).  If $c > rb$, then the safe level ${c \over r} > b$, and it is optimal {\it not} to buy insurance when $b \le W_t < {c \over r}$.  Thus, we can rewrite (2.5) as
$$
\left\{
\eqalign{
&\la \left(\Phi - {\bf 1}_{\{w \ge \wb \}}\right) = \left(rw - c - h(b - w) \, {\bf 1}_{\{ \wb \le w \le \min(\ws, b) \}} \right) \Phi_w + \max_\pi \left[ (\mu - r) \pi \, \Phi_w + {1 \over 2} \sigma^2 \pi^2 \Phi_{ww} \right], \cr
&\Phi(0) = 0, \quad \Phi(\ws) = 1.
}
\right.
\eqno(2.9)
$$}

\rem{2.5} {The problem of maximizing the probability of reaching the bequest goal scales jointly with wealth $w$, the bequest goal $b$, and the rate of consumption $c$.  Specifically, if we write $\phi = \phi(w; b, c)$, $\wb = \wb(b, c)$, and $\pi^* = \pi^*(w; b, c)$ to denote the dependence of $\phi$ and the optimal strategies on $w$, $b$, and $c$, then the following relationships hold.  For any constant $k > 0$,
$$
\phi(kw; kb, kc) = \phi(w; b, c),
\eqno(2.10)
$$
$$
\wb(kb, kc) = k \, \wb(b, c),
\eqno(2.11)
$$
and
$$
\pi^*(kw; kb, kc) = k \, \pi^*(w; b, c).
\eqno(2.12)
$$
The reader will observe these relationships in the solutions in Sections 3 through 5.

Because of this scaling, we could set, say, $b = 1$, solve for $\phi$, $\wb$, and $\pi^*$ with $b = 1$, then obtain the more general quantities as follows: $\phi(w; b, c) = \phi(w/b; 1, c/b)$, $\wb(b, c) = b \, \wb(1, c)$, and $\pi^*(w; b, c) = b \, \pi^*(w/b; 1, c/b)$.  Alternatively, we could set $c = 1$ and proceed similarly.  We choose not to set $b = 1$ or $c = 1$ because we want to allow $b \to 0$ and observe that the solution approaches the one for minimizing the probability of lifetime ruin, and we want to keep $c$ general in order to compare the optimal investment strategy to other optimal investment strategies for which $c$ plays a role.  Furthermore, setting $b = 1$ or $c = 1$ does not simplify the mathematical analysis, except to remove one parameter.  \qed}

\rem{2.6}  {Bayraktar and Young (2009) considered the problem of minimizing shortfall at death without life insurance in the market; they found that the optimal amount to invest in the risky asset is greater than the optimal amount when minimizing lifetime shortfall.  If we were to add life insurance to the market when minimizing $\E^w[(b - W_{\td \wedge \tauo})_+]$ (again, with the understanding that the game ends if wealth reaches $0$, or equivalently, $0$ is an absorbing state for the wealth process), then because the price of life insurance and the shortfall target are (piecewise) linear in the size of the death benefit, if it is optimal to buy any amount of life insurance, it will be optimal to buy $b - w$.  Furthermore, we anticipate that the corresponding value function, $V$, will be decreasing and convex; if so, the optimal life insurance purchasing strategy will be given by (2.7) for $\wb = \inf \{w \ge 0: \la + h \, V_w(w) \ge 0 \} \wedge b$, and $V$ will solve the following BVP:
$$
\left\{
\eqalign{
&\la (V - (b - w)_+) = (rw - c) V_w + \min_\pi \left[ (\mu - r) \pi \, V_w + {1 \over 2} \sigma^2 \pi^2 V_{ww} \right] \cr
& \qquad \qquad \qquad \qquad \; - (b - w)(\la + h V_w) \, {\bf 1}_{\{ \wb \le w \le \min(\ws, b) \}}, \cr
&V(0) = b, \quad V(\ws) = 0.
}
\right.
$$
Thus, even though $V$ measures the magnitude of the shortfall, and not just whether there is shortfall, the optimal insurance purchasing strategy is of the same form as for the bequest goal problem solved in this paper. \qed}

In the following sections, we use Lemma 2.1 to calculate $\phi$.  The solution differs depending on whether $c = 0$, $0 < c \le rb$, or $c > rb$, so we split the problem into those three cases in the next three sections.  Specifically, in Section 3, we consider the case for which $c = 0$ and explicitly determine $\phi$.  In Sections 4 and 5, we consider the cases for which $0 < c \le rb$ and $c > rb$, respectively, and express $\phi$ through its convex Legendre transform (a technique employed in Bayraktar and Young (2007, 2015)), except for the special case in Section 4.1, for which we can write $\phi$ explicitly.

\sect{3. The case for which $c = 0$}

Bayraktar et al.\ (2014, Section 3.1) computed the maximum probability of reaching the bequest goal in the case for which $c = 0$, with no risky asset in the market; let $\phid$ denote the maximum probability in this deterministic case.  In that work, we found that if $r \ge \la$, then the individual does not buy insurance until her wealth reaches the safe level $\ws = {hb \over r + h}$.  At the other extreme, if $r + h \le \la$, then for all levels of wealth, the individual buys term life insurance of $b - w$ until death or ruin because insurance is cheap enough and the riskless return is low enough relative to the hazard rate.

Otherwise, if $r < \la < r + h$, then for wealth less than a certain value ($w^*$ in that paper), the individual buys term life insurance of $b - w$ until death or ruin, while for wealth greater than $w^*$, she does not buy insurance until her wealth reaches the safe level.  We showed that $\phid$ equals
$$
\phid(w) =
\cases{1 - \left( 1 - {w \over \ws} \right)^{{\la \over r +h}}, &if $0 \le w < w^*$, \cr \cr
\left( {w \over \ws} \right)^{{\la \over r}}, &if $w^* \le w \le \ws = {hb \over r + h}$,
}
\eqno(3.1)
$$
in which $w^* = 0$ if $r \ge \la$; otherwise, if $r < \la$, $0 < w^* < \ws$ is the unique value that makes $\phid$ continuous.  In (3.1), $1 - \left( 1 - {w \over \ws} \right)^{{\la \over r +h}}$ is the probability of reaching the bequest goal if the individual buys term life insurance of $b - w$ until she dies or ruins, and $\left( {w \over \ws} \right)^{{\la \over r}}$ is the probability of reaching the bequest goal if the individual does not buy insurance until her wealth reaches the safe level.  When there is no risky asset, the individual buys life insurance at low wealth levels when $\la > r$ because the probability that her wealth will reach the safe level through saving is too small relative to her probability of dying.

When the financial market includes a risky asset, the corresponding strategy for buying term life insurance is turned upside down.  Specifically, for wealth less than the {\it buy level} $\wb$ (``b'' for buy), the individual does {\it not} buy life insurance, while for wealth greater than the buy level $\wb$, she buys insurance of $b - w$.  Because of the existence of a risky asset, when wealth is low, the individual wants to keep her wealth as unconstrained as possible to invest in the risky asset so that she can reach the buy level.

In the no-risky-asset case, there is only one uncertainty, the time of death, so the individual essentially compares the probability of surviving long enough to reach the safe level (while not buying insurance) with the probability of dying before hitting $0$ (while buying insurance).  When there is a risky asset in the market, there is additional uncertainty, the uncertainty in investment returns, and the individual will gamble on that uncertainty to help meet her bequest goal.

Recall that we observed a similar phenomenon when maximizing expected utility of lifetime consumption plus utility of wealth at death.  Pliska and Ye (2007) have no risky asset in their financial market, and the optimal strategy for purchasing insurance decreases with increasing wealth.  The analog in the bequest-goal setting is that, when there is no risky asset and $r < \la < r + h$, the individual only purchases insurance when wealth is small enough.  By contrast, Richard (1975) includes a risky asset in his financial market, and the optimal strategy for purchasing insurance increases with wealth. The analog in the bequest-goal setting is that, when there is a risky asset, the individual only purchases insurance when wealth is large enough. 

In the following proposition, we present $\phi$ and the corresponding optimal strategies for purchasing life insurance and investing in the risky asset.  We omit the proof because it is a straightforward application of Lemma 2.1.

\thm{3.1}  {If $c = 0$, the maximum probability of reaching the bequest goal equals
$$
\phi(w) = 
\cases{
{p(1 - q) \over p - q} \left( {w \over \wb} \right)^q, &if $0 \le w < \wb,$ \cr \cr
1 - {q(p - 1) \over p - q} \left( {\ws - w \over \ws - \wb} \right)^p, &if $\wb \le w \le \ws = {hb \over r+ h},$
}
\eqno(3.2)
$$
in which
$$
q = {1 \over 2r} \left[ (r + \la + m) - \sqrt{(r + \la + m)^2 - 4 r \la} \right] \in (0, 1),
\eqno(3.3)
$$
$$
p = {1 \over 2(r + h)} \left[ (r + h + \la + m) + \sqrt{(r + h + \la + m)^2 - 4(r + h) \la} \right] > 1,
\eqno(3.4)
$$
$$
m = {1 \over 2} \left( {\mu - r \over \sig} \right)^2,
\eqno(3.5)
$$
and
$$
\wb = {1 - q \over p - q} \, \ws \; \in (0, \ws).
\eqno(3.6)
$$
\medskip
When wealth equals $w$, the optimal amount of instantaneous term life insurance equals
$$
D^*(w) = (b - w) \, {\bf 1}_{\{ \wb \le w \le \ws \}},
\eqno(3.7)
$$
and the optimal amount invested in the risky asset equals
$$
\pi^*(w) = 
\cases{
{\mu - r \over \sig^2} \, {w \over 1 - q}, &if $0 < w < \wb,$ \cr \cr
{\mu - r \over \sig^2} \, {\ws - w \over p - 1}, &if $\wb \le w \le \ws$.
}
\eqno(3.8)
$$}

We present the following corollary of Theorem 3.1, in which we give the process of optimally controlled wealth for wealth greater than $\wb$ and observe that wealth never reaches the safe level $\ws$ nor does ruin occur.

\cor{3.2} {If $c = 0$, then the optimally controlled wealth process follows the dynamics
$$
d W^*_t = \cases{
 W^*_t   \left[ \left( r + {2m \over 1 - q} \right) dt + {\mu - r \over \sig} \, {1 \over 1- q} \, dB_t \right], &if $W^*_t < \wb$, \cr \cr
\left( \ws - W^*_t \right)  \left[ \left( {2m \over p - 1} - (r + h) \right) dt + {\mu - r \over \sig} \, {1 \over p - 1} \, dB_t \right], &if $W^*_t > \wb$.
}
$$
Thus, $0 < W^*_t < \ws$ almost surely, for all $t \ge 0$, if $W_0 = w \in (0, \ws)$.  \qed}

\rem{3.1} {Optimally controlled wealth never reaches the safe level because $\ws - W^*$ behaves like geometric Brownian motion for wealth near $\ws$.  Young (2004) observed a similar behavior of optimally controlled wealth when minimizing the probability of lifetime ruin under a constant rate of consumption.  Indeed, the investment strategy when wealth lies between $\wb$ and $\ws$ is quite similar to the investment strategy when minimizing the probability of lifetime ruin, in that both strategies decrease linearly towards zero as wealth increases towards the safe level.  Also, note that ruin will not occur when optimally investing and buying life insurance because $W^*$ behaves like geometric Brownian motion near 0. \qed}

\rem{3.2} {As the hazard rate $\la$ increases to $\infty$, one can show that the buy level $\wb$ decreases to $0$. This monotonicity makes sense because as the person is more likely to die, she will be more willing to buy insurance for a fixed premium rate. Furthermore, as $\la \to \infty$, the individual is likely to die ``in the next second,'' so the only way of reaching a bequest goal is to buy life insurance, which is supported by $\wb \to 0$. \qed}

\sect{4. The case for which $0 < c \le rb$}

When the rate of consumption is large enough and when the premium rate for life insurance is low enough, then it is optimal to buy life insurance of $b - w$ for all $0 \le w \le \ws$ (that is, $\wb = 0$), and we have an explicit expression for $\phi$, as in Theorem 3.1.  We present this case in Section 4.1.

Otherwise, if $\wb > 0$, then we cannot write $\phi$ explicitly.  However, we can write it semi-explicitly by using the Legendre transform to obtain an ansatz for $\phi$; see Bayraktar and Young (2007, 2015) for more details on this technique.

\subsect{4.1  Buying life insurance at all levels of wealth}

When the rate of consumption is large enough and when the premium rate is low enough, it is optimal for the individual to buy life insurance at all levels of wealth.  For this reason, we can explicitly solve the maximization problem; see Theorem 4.2 below.  First, we present the following lemma that we use in the statement of Theorem 4.2.  The proof of the lemma is straightforward, so we omit it.

\lem{4.1} {If $h \le  {r \over r + m} \, \la$, then $C_1 \le rb$, in which $C_1$ is defined by
$$
C_1 = hb \left( {(r+h)p \over \la} - 1 \right).
\eqno(4.1)
$$}

\thm{4.2} {If $h \le  {r \over r + m} \, \la$ and if $C_1 \le c \le rb$, in which $C_1 \le rb$ is given in $(4.1)$, then the maximum probability $\phi$ of reaching the bequest goal equals
$$
\phi(w) = 1 - \left( 1 - {w \over \ws} \right)^p,  \quad 0 \le w \le \ws = {c + hb \over r +h},
\eqno(4.2)
$$
in which $p$ is given in $(3.4)$.  When wealth equals $w \in (0, \ws]$, the optimal amount of instantaneous term life insurance equals
$$
D^*(w) = b - w,
\eqno(4.3)
$$
and the optimal amount invested in the risky asset equals
$$
\pi^*(w) = {\mu - r \over \sig^2} \, {\ws - w \over p - 1}.
\eqno(4.4)
$$}

\pf We use Lemma 2.1 to prove this proposition.  Because $p > 1$, the function $\phi$ in (4.2) is such that $\phi_w > 0$ and $\phi_{ww} < 0$ in $(0, \ws)$.  It clearly satisfies the boundary conditions of (2.5), and one can show that it satisfies the differential equation in (2.5) with $D = b - w$.  As indicated in (2.8), to prove $D = b - w$ is optimal, we must prove
$$
\la - h (b - w) \phi_w(w) \ge 0,
$$
for all $0 \le w \le \ws$, or equivalently,
$$
\la - h(b - w) \, {p \over \ws} \left( 1 - {w \over \ws} \right)^{p-1} \ge 0.
\eqno(4.5)
$$
Inequality (4.5) holds at $w = 0$ if and only if
$$
c \ge C_1.
$$
Furthermore, the left side of inequality (4.5) increases with respect to $w$; thus, if $C_1 \le c \le rb$, then (4.5) holds on $[0, \ws]$.  Finally, the optimal investment strategy in (4.4) follows from (2.6) and (4.2).  \qed

\rem{4.1} {At first blush, it seems as if buying life insurance when consumption is large and when wealth is near zero would be more likely to lead to ruin than if the individual were not to buy life insurance when wealth is near zero.  However, the key is that insurance is inexpensive relative to the rate of dying, $h \le {r \over r + m} \, \la$.  Recall that the individual only wins the game if her wealth at death is at least equal to $b$, so if the consumption rate is great enough ($C_1 \le c \le rb$), the negative tug on wealth from $-c$ means that she is better off buying life insurance at all levels of wealth rather than waiting until wealth reaches some higher level and then buying life insurance.  \qed}

\rem{4.2} {From (4.4), we deduce that the second expression for optimally controlled wealth in Corollary 3.2  also holds for all $W^*_t > 0$, when $C_1 \le c \le rb$.  Thus, $\ws - W^*$ behaves like geometric Brownian motion for wealth near $\ws$, so wealth never reaches the safe level $\ws$.  \qed}

\subsect{4.2  Buying life insurance only when wealth is large enough}

When $0 < c \le rb$ and $c < C_1$, with $C_1$ given in (4.1), it is optimal to buy life insurance only when wealth is larger than some positive level, that is, $\wb > 0$, as in the case for which $c = 0$.  To obtain an ansatz for $\phi$, we (1) hypothesize that $\phi$ solves (2.9) with $\wb > 0$, $\phi_w > 0$, and $\phi_{ww} < 0$, (2) formally define $\phi$'s convex Legendre transform and determine its free-boundary problem, (3) solve this dual free-boundary problem, and (4) compute the concave Legendre transform of that solution.  Bayraktar and Young (2007, Section 4.1) follow these steps to minimize the expectation of a non-increasing, non-negative function of minimum wealth in a Black-Scholes market.  Also, Bayraktar and Young (2015) perform steps (3) and (4) when maximizing the probability of reaching the bequest goal without life insurance in the market.  They include those steps because, when $0 < c < rb$, the convex Legendre transform is the value function of an optimal stopping problem.

We omit the details of steps (1)-(4) because they are similar to those found in Bayraktar and Young (2007, Section 4.1).  Instead, in the following theorem, we present the candidate for $\phi$ thus obtained and verify directly that it satisfies the conditions of Lemma 2.1.

\thm{4.3} {If $0 < c \le rb$ and if $c < C_1$, then the maximum probability of reaching the bequest goal equals
$$
\phi(w) =
\cases{
{c \over r} \, {(\al_1 - 1)(1 - \al_2) \over \al_1 - \al_2} \left[ - \left( {y \over \yo} \right)^{\al_1} + \left( {y \over \yo} \right)^{\al_2} \right] \yo, &if $0 \le w < \wb = b - {\la \over h} {1 \over \yb},$ \cr \cr
1 - {\la \over h p} \, {\ws - \wb \over b - \wb} \left( {\ws - w \over \ws - \wb} \right)^p, &if $\wb \le w \le \ws = {c + hb \over r+ h},$
}
\eqno(4.6)
$$
in which
$$
\eqalign{
\be_1 &= {1 \over 2m} \left[ (r + h - \la + m) + \sqrt{(r +h - \la + m)^2 + 4 m \la} \right] > \al_1, \cr
\al_1 &= {1 \over 2m} \left[ (r - \la + m) + \sqrt{(r - \la + m)^2 + 4 m \la} \right] > 1, \cr
\al_2 &= {1 \over 2m} \left[ (r - \la + m) - \sqrt{(r - \la + m)^2 + 4 m \la} \right] < 0,}
\eqno(4.7)
$$
with $p = {\be_1 \over \be_1 - 1}$ and $m$ given in $(3.4)$ and $(3.5)$, respectively.  The parameter $\yb > 0$ is given by
$$
{\la \over h} \, {1 \over \yb} = b + {c \over r} \left[ {\al_1 (1 - \al_2) \over \al_1 - \al_2} \, \ybo^{\al_1 - 1} + {\al_2 (\al_1 - 1) \over \al_1 - \al_2} \, \ybo^{\al_2 - 1} - 1 \right] ,
\eqno(4.8)
$$
in which $\ybo \in (0, 1)$ uniquely solves
$$
{c \over r} \left[{\al_1 (1 - \al_2) (\be_1 - \al_1) \over \al_1 - \al_2} \, \ybo^{\al_1 - 1} + {\al_2 (\al_1 - 1) (\be_1 - \al_2) \over \al_1 - \al_2} \, \ybo^{\al_2 - 1} \right] = (\be_1 - 1) \left( {c \over r} - \ws \right),
\eqno(4.9)
$$
and the parameter $\yo > \yb$ is given by
$$
\yo = {\yb \over \ybo}.
\eqno(4.10)
$$
In the first expression of $(4.6)$, for a given $w \in [0, \wb)$, $y \in (\yb, \yo]$ uniquely solves
$$
{c \over r} \left[{\al_1(1 - \al_2) \over \al_1 - \al_2} \, \left( {y \over \yo} \right)^{\al_1 - 1} + {\al_2(\al_1 - 1) \over \al_1 - \al_2} \, \left( {y \over \yo} \right)^{\al_2 - 1} \right] = {c \over r} - w.
\eqno(4.11)
$$
\indent When wealth equals $w \in (0, \ws]$, the optimal amount of instantaneous term life insurance equals
$$
D^*(w) = (b - w) \, {\bf 1}_{\{\wb \le w \le \ws\}},
\eqno(4.12)
$$
and the optimal amount invested in the risky asset equals
$$
\pi^*(w) = 
\cases{
{\mu - r \over \sig^2} \, {c \over r} \, {(\al_1 - 1)(1 - \al_2) \over \al_1 - \al_2} \left[ \al_1 \left( {y \over \yo} \right)^{\al_1 - 1} - \al_2 \left( {y \over \yo} \right)^{\al_2 - 1} \right], &if $0 < w < \wb,$ \cr \cr
{\mu - r \over \sig^2} \, {\ws - w \over p - 1}, &if $\wb \le w \le \ws$.
}
\eqno(4.13)
$$}

\pf First, note that there exists a unique solution $\ybo \in (0, 1)$ of (4.9).  Indeed, the left side of (4.9) increases with respect to $\ybo$, and as $\ybo$ approaches $0+$, the left side of (4.9) approaches $-\infty$. When $\ybo = 1$, the left side becomes
$$
{c \over r } \left( \be_1 - {r + m \over m} \right),
$$
which is greater than the right side if and only if $c$ is less than
$$
hb \; {m \over (r + h) p - (r + h + m)} \, ,
$$
which equals $C_1$.  Thus, because $c < C_1$, there exists a unique solution in $(0, 1)$ of (4.9).

Second, we prove that $\yb > 0$.  It is straightforward to show that $\ybo$ increases with $c$, which implies that the right side of (4.8) increases with $c$. So, it is enough to show that the right side of (4.8) is positive as $c \to 0+$.  From (4.9), we deduce that
$$
\lim_{c \to 0+} {c \over r} {\al_2 (\al_1 - 1) (\be_1 - \al_2) \over \al_1 - \al_2} \, \ybo^{\al_2 - 1} = - (\be_1 - 1) {hb \over r + h}.
$$
Thus, the limit of the right side of (4.8), as $c \to 0+$, equals
$$
\eqalign{
b - {\be_1 - 1 \over \be_1 - \al_2} \, {hb \over r + h} > 0.}
$$

Third, we prove that $\wb$ defined by $\wb = b - {\la \over h} \, {1 \over \yb}$ is positive.  In terms of $\ybo$, we write
$$
\wb =  {c \over r} \left[ 1 - {\al_1 (1 - \al_2) \over \al_1 - \al_2} \, \ybo^{\al_1 - 1} - {\al_2 (\al_1 - 1) \over \al_1 - \al_2} \, \ybo^{\al_2 - 1} \right],
\eqno(4.14)
$$
and the quantity in the square brackets decreases with $c$.  We have two cases to consider $h \le {r \over r+m} \, \la$ and $h > {r \over r+m} \, \la$.  If $h \le {r \over r+m} \, \la$, then $C_1 \le rb$, so it is enough to show that $\wb \ge 0$ when $c = C_1$.  From the above discussion, we know that if $c = C_1$, then $\ybo = 1$; thus, (4.14) gives
$$
\wb \big|_{c = C_1} = 0.
$$
If $h > {r \over r+m} \, \la$, then $C_1 > rb$, so it is enough to show that $\wb > 0$ when $c = rb$.  If $c = rb$, then
$$
\ybo = \left( - {\al_2(\al_1 - 1)(\be_1 - \al_2) \over \al_1(1 - \al_2)(\be_1 - \al_1)} \right)^{1 \over \al_1 - \al_2},
$$
and (4.14) gives
$$
\wb \big|_{c = rb} = {c \over r} \left[ 1 - \left( {\al_1(1 - \al_2) \over \be_1 - \al_2} \right)^{{1 - \al_2 \over \al_1 - \al_2}} \left( - {\al_2(\al_1 - 1) \over \be_1 - \al_1} \right)^{{\al_1 - 1 \over \al_1 - \al_2}} \right],
$$
which increases with $h$ because $\be_1$ increases with $h$.  Thus, it is enough to show that $\wb \big|_{c = rb} \ge 0$ when $h = {r \over r+m} \, \la$.  If $h = {r \over r+m} \, \la$, then $rb = C_1$, and $\wb \big|_{c = rb} = 0$.

Fourth, we prove that $\wb < \ws$.  From (4.9) and (4.14), we deduce that $\wb < \ws$ if and only if
$$
{\al_1 (1 - \al_2) (\be_1 - \al_1) \over \al_1 - \al_2} \, \ybo^{\al_1 - 1} + {\al_2 (\al_1 - 1) (\be_1 - \al_2) \over \al_1 - \al_2} \, \ybo^{\al_2 - 1} <  {\al_1 (1 - \al_2) \over \al_1 - \al_2} \, \ybo^{\al_1 - 1} + {\al_2 (\al_1 - 1) \over \al_1 - \al_2} \, \ybo^{\al_2 - 1},
$$
which is equivalent to $\al_2 \, \ybo^{\al_2 -1} < \al_1 \, \ybo^{\al_1 - 1}$, which holds because $\al_2 < 0 < 1 < \al_1$.

Finally, we use Lemma 2.1 to show that the expression in (4.6) equals the maximum probability of reaching the bequest goal.  When $w = 0$, (4.11) implies that $y = \yo$, so the first expression in (4.6) equals $0$.  Clearly, when $w = \ws$, the second expression in (4.6) equals $1$.  For $0 \le w < \wb$,
$$
\phi_w(w) = y > 0,
\eqno(4.15)
$$
and
$$
\phi_{ww}(w) = y \, \left( {c \over r} \, {(\al_1 - 1)(1 - \al_2) \over \al_1 - \al_2} \left[ - \al_1 \left( {y \over \yo} \right)^{\al_1 - 1} + \al_2 \left( {y \over \yo} \right)^{\al_2 - 1} \right] \right)^{-1} < 0.
\eqno(4.16)
$$
Using (4.15) and (4.16), one can show that the first expression in (4.6) satisfies the differential equation in (2.9).  It is straightforward to show that the second expression in (4.6) also satisfies the differential equation in (2.9) in $(\wb, \ws)$, with $\phi_w > 0$ and $\phi_{ww} < 0$ in that interval.

We next show that $\phi$ given in (4.6) is $\C^2$ at $w = \wb$.  To that end, note that when $w = \wb$, the solution of (4.11) is $y = \yb$; thus, from (4.15),
$$
\phi_w(\wb-) = \yb,
$$
and from the second expression in (4.6),
$$
\phi_w(\wb+) = {\la \over h} \, {1 \over b - \wb} = \yb.
$$
Next, from (4.16),
$$
\phi_{ww}(\wb-) = \yb \, \left( {c \over r} \, {(\al_1 - 1)(1 - \al_2) \over \al_1 - \al_2} \left( - \al_1 \, \ybo^{\al_1 - 1} + \al_2 \, \ybo^{\al_2 - 1} \right) \right)^{-1},
$$
and the second expression in (4.6) gives
$$
\phi_{ww}(\wb+) = - {\la \over h} \, {p - 1 \over (b - \wb)(\ws - \wb)} = - \yb \, {p - 1 \over \ws - \wb}.
$$
The equality of these expressions for $\phi_{ww}(\wb-)$ and $\phi_{ww}(\wb+)$ follows from (4.9) after substituting for $\wb$ via (4.14).  Next, from the differential equation in (2.9),
$$
\la \phi(\wb-) = (r \wb - c) \phi_w(\wb) - m {\phi_w^2(\wb) \over \phi_{ww}(\wb)},
$$
and
$$
\la \phi(\wb+) = (r \wb - c) \phi_w(\wb) - m {\phi_w^2(\wb) \over \phi_{ww}(\wb)} + (\la - h(b - \wb)\phi_w(\wb)).
$$
From $\phi_w(\wb) = \yb$ and $\wb = b - {\la \over h} \, {1 \over \yb}$, we deduce that $\phi(\wb-) = \phi(\wb+)$.  Thus, we have shown that $\phi$ is $\C^2$ at $w = \wb$. Furthermore, we have shown that the expression given in (4.6) satisfies the conditions of Lemma 2.1 and, thus, equals the maximum probability of reaching the bequest goal.

The optimal strategies in (4.12) and (4.13) follow from (2.6) and (2.7), respectively.  \qed

\rem{4.3} {When it is optimal not to buy life insurance, that is, when $0 < w < \wb$, then the optimal amount invested in the risky asset in (4.13) is {\it independent} both of the bequest goal $b$ and of the premium rate $h$, a surprising result.  Indeed, the solution ${y \over \yo}$ of (4.11) is independent of $b$ and $h$; thus, the first expression in (4.13) is independent of $b$ and $h$.  Also, note that this independence holds when $c = 0$; see (3.8) in Theorem 3.1.

Moreover, we will see in Proposition 6.4 that, when $0 < w < \wb$, $\pi^*(w)$ is identical to the amount invested in the risky asset when life insurance is not available in the market.  Thus, one might interpret this by saying that, when $0 < w < \wb$, the individual is investing to attain any level of wealth greater than current wealth. 

When it is optimal to buy life insurance, that is, when $\wb \le w \le \ws$, the optimal investment strategy in (4.13) is {\it identical} to the one given in (4.4) in Theorem 4.2.  Also, note that the relationships in (2.10) through (2.12) hold for the solution in Theorem 4.3, as expected.  \qed}

\rem{4.4} {From the expression for $\pi^*(w)$ in (4.13) when $\wb \le w \le \ws$, we deduce that the expression for optimally controlled wealth in Corollary 3.2 when $W^*_t > \wb$ also holds when $0 < c \le rb$.  Thus, wealth never reaches the safe level.  \qed}

We expected $\wb$ to decrease monotonically from ${1 - q \over p - q} \, {hb \over r + h}$ as $c$ increases because if $h \le {r \over r + m} \, \la$, then $\wb \big|_{c = C_1} = 0$.  However, the following corollary shows that $\wb$ first {\it increases} and then decreases as $c$ increases, a surprising result.

\cor{4.4} {Suppose $h \le {r \over r + m} \, \la$, which implies $C_1 \le rb$.  As $c$ increases from $0$ to $C_1$, the buy level $\wb$ first increases from the expression given in $(3.6)$ and then decreases to $0$.  Moreover, if $h > {r \over r + m} \, \la$, then the buy level $\wb$ increases at $c = 0+$.}

\pf By differentiating (4.14) with respect to $c$ and by substituting for ${\partial \ybo \over \partial c}$, which we obtain by differentiating (4.9) with respect to $c$, we get
$$
\eqalign{
{\partial \wb \over \partial c} &= {1 \over r} \left[ 1 - {\al_1 (1 - \al_2) \over \al_1 - \al_2} \, \ybo^{\al_1 - 1} - {\al_2 (\al_1 - 1) \over \al_1 - \al_2} \, \ybo^{\al_2 - 1} \right] \cr
&\quad - (\be_1 - 1) \, {hb \over c(r + h)} \, {\al_1 \, \ybo^{\al_1 - 1} - \al_2 \, \ybo^{\al_2 - 1} \over {\al_1(\be_1 - \al_1) \ybo^{\al_1 - 1} - \al_2 (\be_1 - \al_2) \ybo^{\al_2 - 1}}}.
}
$$
After substituting for $(\be_1 - 1) \, {hb \over c(r + h)}$ from (4.9) and simplifying, we get
$$
\eqalign{
{\partial \wb \over \partial c} &\propto \al_1 \left( {r \over r + h} \, \be_1 - \al_1 + {h \over r + h} \right) \ybo^{1 - \al_2} - \al_2 \left( {r \over r + h} \, \be_1 - \al_2 + {h \over r + h} \right) \ybo^{1- \al_1} + \al_1 \al_2(\al_1 - \al_2).
}
\eqno(4.17)
$$
Define a function $k$ of $y \in (0, 1)$ by the right side of (4.17).  As $c \to 0+$, $\ybo \to 0+$, and $\lim_{y \to 0+} k(y) = +\infty$.  As $c \to C_1-$, $\wb \to 0+$, $\ybo \to 1-$, and
$$
\lim_{y \to 1-} k(y) \propto m \be_1 - (r + h+ m) < 0.
$$
The derivative of $k$ is proportional to
$$
k'(y) \propto \al_1 (1 - \al_2) \left( {r \over r + h} \, \be_1 - \al_1 + {h \over r + h} \right) y^{\al_1 - \al_2} + \al_2 (\al_1 - 1) \left( {r \over r + h} \, \be_1 - \al_2 + {h \over r + h} \right),
$$
which is negative for all $y \in (0, 1)$ if $k'(1) \le 0$, or equivalently, if
$$
{r \be_1 + h \over r + h} \le {r + m \over m}.
\eqno(4.18)
$$
It is straightforward to show that the left side of inequality (4.18) increases with $h$, and $\lim_{h \to \infty} {r \be_1 + h \over r + h} = {r + m \over m}$.  Thus, (4.18) holds strictly for all $h \le {r \over r + m} \, \la$, and we deduce that ${\partial \wb \over \partial c}$ switches from positive to negative as $c$ increases from $0$ to $C_1$.  Moreover, ${\partial \wb \over \partial c} \big|_{c = 0+}$ is positive, independent of the magnitude of $h$.   \qed

\rem{4.5} {We were surprised about the lack of monotonicity of $\wb$ with respect to $c$, and we initially thought it might be due to the fact that $\ws$ also increases with $c$.  That is, perhaps the difference, $\ws - \wb$, increases with $c$.  However, in work parallel to the proof of Corollary 4.4, one can show that $\ws - \wb$ first decreases and then increases as $c$ increases.  Specifically, as $c$ initially increases from $0+$, $\wb$ increases more quickly than $\ws$ does. \qed}

\sect{5.  The case for which $c > rb$}

When the safe level $\ws = {c \over r} > b$, it is optimal not to buy life insurance when $w \ge b$.  When the rate of consumption is large enough or when the premium rate for life insurance is low enough, it is optimal to buy life insurance with death benefit $b - w$ for all $0 < w < b$; we present this case in Section 5.1.  Otherwise, it is optimal to buy life insurance only for wealth lying between $\wb > 0$ and $b$; we present this case in Section 5.2.

\subsect{5.1 Buying life insurance at all levels of wealth less than the bequest goal}

When the rate of consumption is large enough or when the premium rate is low enough, it is optimal to buy life insurance at all levels of wealth less than the bequest goal.  Unlike the problem in Section 4.1, we do not have an explicit solution for $\phi$ because it is optimal not to buy life insurance for wealth between $b$ and ${c \over r}$.  Thus, we rely on $\phi$'s convex Legendre transform, as in Section 4.2, to obtain an ansatz for $\phi$.  Again, we omit the details because they are similar to those in Bayraktar and Young (2007, 2015). In Theorem 5.2 below, we present the candidate for $\phi$ and verify directly that it satisfies the conditions of Lemma 2.1. First, we prove a useful lemma.

\lem{5.1} {Define the function $g$ by
$$
g(\be) = r - (r + h) \, {\be \over \al_1} + h \be,
\eqno(5.1)
$$
in which $\al_1$ is given in $(4.7)$.  Then,
$$
g(\be_1) > 0 \hbox{ and } \, g(\be_2) > 0,
$$
in which $\be_1$ is given in $(4.7)$ and
$$
\be_2 = {1 \over 2m} \left[ (r + h - \la + m) - \sqrt{(r +h - \la + m)^2 + 4 m \la} \right] < 0.
\eqno(5.2)
$$}

\pf First, $g(\be_1) > 0$ if and only if $r \al_1 + ((\al_1 - 1)h - r) \be_1 > 0$.  When $h = 0$, the latter inequality holds with equality because $\be_1 \big|_{h=0} = \al_1$.  Thus, if we show that $r \al_1 + ((\al_1 - 1)h - r) \be_1$ increases with $h$, then we are done.
$$
\eqalign{
{\partial \over \partial h} (r \al_1 + ((\al_1 - 1)h - r) \be_1) &= (\al_1 - 1) \be_1 +  {((\al_1 - 1)h - r) \be_1 \over \sqrt{(r +h - \la + m)^2 + 4 m \la}} \cr
&\propto (\al_1 - 1)(h + \sqrt{(r +h - \la + m)^2 + 4 m \la})  - r,
}
$$
which is positive if it is non-negative at $h = 0$ because $h + \sqrt{(r +h - \la + m)^2 + 4 m \la}$ increases with $h$.  To show that
$$
(\al_1 - 1)\sqrt{(r - \la + m)^2 + 4 m \la}  - r \ge 0,
$$
use $\al_1 - 1 = {1 \over \po - 1}$, in which $\po = p \big|_{h=0}$, with $p$ given in (3.4), to obtain the equivalent inequality
$$
\sqrt{(r + \la + m)^2 - 4 r \la} - r(\po - 1) \ge 0,
$$
or
$$
\sqrt{(r + \la + m)^2 - 4 r \la} \ge (r + \la + m) - 2r.
$$
The latter inequality is straightforward to demonstrate; thus, we have shown $g(\be_1) > 0$.

Next, to prove $g(\be_2) > 0$, it is enough to show that $g(\be_2)$ decreases with $h$ and that \break $\lim_{h \to \infty} g(\be_2) \ge 0$.
$$
{\partial \over \partial h} \left(r - (r + h) \, {\be_2 \over \al_1} + h \be_2 \right) \propto (\al_1 - 1)(h - \sqrt{(r +h - \la + m)^2 + 4 m \la})  - r,
$$
which is negative if it is non-positive as $h \to \infty$, because $h - \sqrt{(r +h - \la + m)^2 + 4 m \la}$ increases with $h$.  The limit of $h - \sqrt{(r +h - \la + m)^2 + 4 m \la}$ as $h \to \infty$ equals $-(r - \la + m)$; thus,
$$
\lim_{h \to \infty} (\al_1 - 1)(h - \sqrt{(r +h - \la + m)^2 + 4 m \la})  - r = - (\al_1 - 1) (r - \la + m) - r,
$$
which one can show is non-positive.  Thus, we have shown that $g(\be_2)$ decreases with $h$.

To finish proving $g(\be_2) > 0$, we observe that, because $\lim_{h \to \infty} \be_2 = 0$ and $\lim_{h \to \infty} h \be_2 = - \la$,
$$
\lim_{h \to \infty} \left(r - (r + h) \, {\be_2 \over \al_1} + h \be_2 \right) = r + {\al_1 - 1 \over \al_1} (-\la) \propto r \po - \la,
$$
in which $\po = p|_{h = 0}$.  It is easy to show that $r \po - \la > 0$; thus, we have proved $g(\be_2) > 0$.  \qed

\thm{5.2} {Suppose $c > rb$, and suppose one of the following two conditions holds: \hfill \break
\noi$(1)$ $h \le {r \over r + m} \, \la$, or \hfill \break
\noi$(2)$  $h > {r \over r + m} \, \la$ and $c \ge C_2$, in which $C_2 > rb$ uniquely solves
$$
{c - rb \over r(r+h)} \left[ {c - rb \over r(r+h)} {g(\be_2) \over {hb \over \la} \, \be_2 + {c + hb \over r + h} (1 - \be_2)} \right]^{{1 - \be_2 \over \be_1 - 1}} = {{hb \over \la} \, \be_1 - {c + hb \over r + h} (\be_1 - 1) \over g(\be_1)}.
\eqno(5.3)
$$
Then, the maximum probability of reaching the bequest goal equals
$$
\phi(w) =
\cases{1 - {c - rb \over r(r + h)} \, \left[{\be_1 - 1 \over \be_1 - \be_2} \, g(\be_2) \left( {y \over \yg} \right)^{\be_1} + {1 - \be_2 \over \be_1 - \be_2} \, g(\be_1) \left( {y \over \yg} \right)^{\be_2} \right] \yg
, &if $0 \le w < b,$ \cr \cr
1 - \left( {c \over r} - b \right) {\yg \over \po} \left( {{c \over r} - w \over {c \over r} - b} \right)^{\po}, &if $b \le w \le \ws = {c  \over r },$
}
\eqno(5.4)
$$
in which $\be_1$, $\al_1$, and $\al_2$ are given in $(4.7)$; $\be_2$, in $(5.2)$; and, $g$, in $(5.1)$.  The parameter $\yo > 0$ is given by
$$
{1 \over \yo} = {c +hb \over r + h} \, {\be_1 - 1 \over \be_1} + {c - rb \over r(r + h)} \, {g(\be_1) \over \be_1} \, \ygo^{1 - \be_2},
\eqno(5.5)
$$
in which $\ygo \in (0, 1)$ uniquely solves
$$
{c - rb \over r(r + h)} \left[ {\be_1 \over \be_1 - \be_2} \, g(\be_2) \, \ygo^{1 - \be_1} - {\be_2 \over \be_1 - \be_2} \, g(\be_1) \, \ygo^{1 - \be_2} \right] = {c + hb \over r + h},
\eqno(5.6)
$$
and $\yg = \yo \ygo$.  Also, $\po = p \big|_{h=0} = {\al_1 \over \al_1 - 1}$, in which $p$ is given in $(3.4)$.  In the first expression of $(5.4)$, for a given $w \in [0, b)$, $y \in (\yg, \yo]$ uniquely solves
$$
{c - rb \over r(r + h)} \, \left[{\be_1 \over \be_1 - \be_2} \, g(\be_2) \left( {y \over \yg} \right)^{\be_1 - 1} - {\be_2 \over \be_1 - \be_2} \, g(\be_1) \left( {y \over \yg} \right)^{\be_2 - 1} \right] = {c + hb \over r + h} - w.
\eqno(5.7)
$$
\indent When wealth equals $w \in (0, \ws]$, the optimal amount of instantaneous term life insurance equals
$$
D^*(w) = (b - w)_+,
\eqno(5.8)
$$
and the optimal amount invested in the risky asset equals
$$
\pi^*(w) = 
\cases{
{\mu - r \over \sig^2} \, {c - rb \over r(r + h)} \left[ {\be_1 (\be_1 - 1) \over \be_1 - \be_2} g(\be_2) \left( {y \over \yg} \right)^{\be_1 - 1} + {\be_2 (1 - \be_2) \over \be_1 - \be_2} g(\be_1) \left( {y \over \yg} \right)^{\be_2 - 1} \right], &if $0 < w < b,$ \cr \cr
{\mu - r \over \sig^2} \, {{c \over r} - w \over \po - 1}, &if $b \le w \le \ws$.
}
\eqno(5.9)
$$}

\pf  We begin by showing that equation (5.6) has a unique solution $\ygo \in (0, 1)$.  Note that the left side approaches $+ \infty$ as $\ygo \to 0+$ because $g(\be_2) > 0$.  Next, when $\ygo = 1$, the left side equals ${c - rb \over r + h}$, which is less than the right side.  Finally, the left side decreases with $\ygo \in (0, 1)$ if and only if
$$
- \be_1 (\be_1 - 1) g(\be_2) - \be_2 (1 - \be_2) g(\be_1) \, \ygo^{\be_1 - \be_2} < 0,
$$
for all $0 < \ygo < 1$, which holds if and only if it holds weakly when $\ygo = 1$ because $g(\be_1) > 0$.  One can show that $- \be_1 (\be_1 - 1) g(\be_2) - \be_2 (1 - \be_2) g(\be_1) \le 0$ is equivalent to $r \po - \la \ge 0$, which is true; see the last line of the proof of Lemma 5.1.  (Recall $\po = p \big|_{h = 0}$.)

Second, we prove that the expression in (5.4) satisfies the BVP in (2.9) with $\wb$ set equal to $0$.  As in the proof of Theorem 4.3, for $0 \le w < b$,
$$
\phi_w(w) = y > 0,
$$
and
$$
\phi_{ww}(w) = -  y \left( {c - rb \over r(r + h)} \left[ {\be_1 (\be_1 - 1) \over \be_1 - \be_2} g(\be_2) \left( {y \over \yg} \right)^{\be_1 - 1} + {\be_2 (1 - \be_2) \over \be_1 - \be_2} g(\be_1) \left( {y \over \yg} \right)^{\be_2 - 1} \right] \right)^{-1}.
$$
The expression for in square brackets increases with $y \in (\yg, \yo]$. Thus, $\phi_{ww} < 0$ for $0 < w < b$ if the expression in square brackets is non-negative when $y = \yg$, which is equivalent to $r \po - \la \ge 0$, which is easy to show.  The remainder of the proof that (5.4) satisfies (2.9) is similar to the corresponding proof of Theorem 4.2, including the proof that $\phi$ in (5.4) is $\C^2$ at $w = b$, so we omit those details in the interest of space.

Third, we prove that $\wb = 0$, in which $\wb$ is defined in (2.8).  Specifically, we show that \hfill \break $\la - h(b - w) \phi_w(w) \ge 0$ for all $0 \le w < b$, with $\phi_w$ determined by the first expression in (5.4).  Because $\phi_{ww} < 0$, this inequality holds for all $0 \le w < b$ if and only if it holds at $w = 0$, or equivalently,
$$
{1 \over \yo} \ge {hb \over \la},
$$
because $\yo = \phi_w(0)$.  Substitute for ${1 \over \yo}$ from the expression in (5.5) to obtain the equivalent inequality
$$
{c - rb \over r(r + h)} \, \ygo^{1 - \be_2} \ge {{hb \over \la} \, \be_1 - {c +hb \over r + h} \, (\be_1 - 1) \over  g(\be_1)}.
\eqno(5.10)
$$
The right side of (5.10) is non-positive exactly when $c \ge C_1$, in which $C_1$ is given in (4.1).  Recall that $C_1 \le rb$ if and only if $h \le {r \over r + m} \, \la$.  Thus, inequality (5.10) holds for all $c > rb$ if $h \le {r \over r + m} \, \la$.

For the remainder of the proof of inequality (5.10), assume that $h > {r \over r + m} \, \la$; then, $C_1 > rb$, and inequality (5.10) automatically holds for all $c \ge C_1$.  Thus, suppose $c \in (rb, C_1)$, so that the right side of (5.10) is positive.  Because the left side of (5.6) decreases with $\ygo$, inequality (5.10) is equivalent to
$$
\eqalign{
{c - rb \over r(r + h)} \, {\be_1 \over \be_1 - \be_2} \, g(\be_2) \, \left[ {{hb \over \la} \, \be_1 - {c +hb \over r + h} \, (\be_1 - 1) \over  {c - rb \over r(r + h)} \, g(\be_1)} \right]^{-{\be_1 - 1 \over 1 - \be_2}} \cr
\quad - {\be_2 \over \be_1 - \be_2} \left( {hb \over \la} \, \be_1 - {c +hb \over r + h} \, (\be_1 - 1) \right) & \ge {c + hb \over r + h}, \cr \cr
\iff \left[ {{hb \over \la} \, \be_1 - {c +hb \over r + h} \, (\be_1 - 1) \over  {c - rb \over r(r + h)} \, g(\be_1)} \right]^{-{\be_1 - 1 \over 1 - \be_2}} &\ge {{hb \over \la} \, \be_2 + {c + hb \over r + h} (1 - \be_2) \over {c - rb \over r(r + h)} \, g(\be_2)}.}
$$
By following an argument similar to the proof of Lemma 5.1, one can show that the numerator of the right side of this inequality is positive, and Lemma 5.1 implies that the denominator is positive.  Thus, this inequality is equivalent to
$$
{c - rb \over r(r+h)} \left[ {c - rb \over r(r+h)} {g(\be_2) \over {hb \over \la} \, \be_2 + {c + hb \over r + h} (1 - \be_2)} \right]^{{1 - \be_2 \over \be_1 - 1}} \ge {{hb \over \la} \, \be_1 - {c + hb \over r + h} (\be_1 - 1) \over g(\be_1)}.
\eqno(5.11)
$$
The left side of (5.11) increases with $c$ and the right side decreases with $c$. It follows that there exists a unique solution $C_2 \in (rb, C_1)$ of (5.3) such that (5.11) holds for all $c \in [C_2, C_1)$.  Recall that (5.10) holds automatically for $c \ge C_1$.  We have shown that, when $h > {r \over r + m} \, \la$, inequality (5.10) holds for all $c \ge C_2$.  Thus, $\wb = 0$.

Finally, the optimal strategies in (5.8) and (5.9) follow from (2.6) and (2.7), respectively.  \qed

Note that the solution in Theorem 5.2 satisfies the relationships in (2.10) through (2.12).  Also, observe that the solution is continuous as the bequest goal approaches $0$, that is, as the bequest goal becomes smaller, then we expect (5.4) and (5.9) to approach 1 minus the minimum probability of lifetime ruin and the corresponding optimal investment strategy, respectively, as obtained in Young (2004).  The following corollary states this result more formally.

\cor{5.3} {The solution given in Theorem $5.5$ is continuous as $b \to 0+$.  In particular, for $0 \le w \le {c \over r}$,
$$
\lim_{b \to 0+} \phi(w) = 1 - \left( 1 - {rw \over c} \right)^{\po},
\eqno(5.12)
$$
and
$$
\lim_{b \to 0+} \pi(w) = {\mu - r \over \sig^2} \, {{c \over r} - w \over \po - 1}.
\eqno(5.13)
$$}

In the next corollary, we show that $\pi^*$ in (5.9) decreases with wealth.

\cor{5.4} {If $c > rb$ and if either condition in Theorem $5.2$ holds, then the optimal amount invested in the risky asset decreases as wealth increases.}

\pf Clearly, $\pi^*(w)$ decreases with wealth when $b \le w \le {c \over r}$.  For $0 < w < b$, differentiate the first expression in (5.9) with respect to $w$ to learn that
$$
{d \pi^*(w) \over dw} \propto {\partial \over \partial w} {y \over \yg}. 
$$
Next, differentiate (5.7) with respect to $w$ to obtain
$$
{\partial \over \partial w} {y \over \yg} \propto - \left[{\be_1(\be_1 - 1) \over \be_1 - \be_2} \, g(\be_2) \left( {y \over \yg} \right)^{\be_1 - 1} + {\be_2(1 - \be_2) \over \be_1 - \be_2} \, g(\be_1) \left( {y \over \yg} \right)^{\be_2 - 1} \right] \propto - \pi^*(w) < 0.
\eqno{\square}
$$

\rem{5.1} {When minimizing the probability of lifetime ruin, the optimal amount invested in the risky asset decreases (linearly) with wealth.  For the problem in this paper, if the individual is buying life insurance, then the bequest goal is covered and the remaining problem is to avoid ruin.  Thus, we expect $\pi^*$ in (5.9) to share properties with the optimal investment strategy for the problem of minimizing the probability of lifetime ruin.  In fact, $\pi^*$ for wealth between $b$ and ${c \over r}$ is {\it identical} to the amount invested in the risky asset when minimizing the probability of lifetime ruin (Young, 2004).  Also, as in Young (2004) and as in the cases considered in Sections 3 and 4, wealth never reaches the safe level.  \qed}

\subsect{5.2  Buying life insurance only when wealth is large enough and less than the bequest goal}

From Theorem 5.2, we know that the remaining case for us to address is $h > {r \over r + m} \, \la $ and $rb < c < C_2$.  Recall that $h > {r \over r + m} \, \la $ is a necessary condition for a solution $C_2 > rb$ of (5.3) to exist.  For this case, based on Theorem 3.1 and Theorem 4.3, we hypothesize that the buy level $\wb$ is positive.   Under this hypothesis, in work not shown here, we solve the free-boundary problem of $\phi$'s convex Legendre transform.  In Theorem 5.7 below, we show that the concave Legendre transform of this (unstated) solution of the free-boundary problem equals the maximum probability of reaching the bequest goal.  First, we prove two useful lemmas.

\lem{5.5} {Define the function $\ell$ by
$$
\ell(\al, \be) = \be - \left( {h \over \la} \, \be + 1 \right) \al.
\eqno(5.14)
$$
Then,
$$
\ell(\al_1, \be_1) < 0, \quad \ell(\al_1, \be_2) < 0, \quad \ell(\al_2, \be_1) > 0, \; \hbox{ and } \; \ell(\al_2, \be_2) < 0, 
$$
in which $\al_1$, $\al_2$, and $\be_1$ are given in $(4.7)$, and $\be_2$ in $(5.2)$.}

\pf We prove these four inequalities in turn.  First, when $h = 0$, $\ell(\al_1, \be_1) = \al_1 - \al_1 = 0$.  Thus, to show that $\ell(\al_1, \be_1) < 0$, it is enough to show that $\ell(\al_1, \be_1)$ decreases with $h$.  To that end,
$$
{\partial \over \partial h} \ell(\al_1, \be_1) \propto \la - \al_1 \left( h + \sqrt{(r + h - \la + m)^2 + 4m \la} \right),
$$
and this expression is negative for all $h > 0$ if it is negative when $h = 0$, which is straightforward to show.

Second, $\lim_{h \to \infty} \ell(\al_1, \be_2) = 0 - \left( {1 \over \la} \, (-\la) + 1 \right) \al_1 = 0$.  Thus, to show that $\ell(\al_1, \be_2) < 0$, it is enough to show that $\ell(\al_1, \be_2)$ increases with $h$.  To that end,
$$
{\partial \over \partial h} \ell(\al_1, \be_2) \propto \la - \al_1 \left( h - \sqrt{(r + h - \la + m)^2 + 4m \la} \right),
$$
and, because this expression decreases with $h$, it is positive for all $h > 0$ if its limit, as $h$ approaches $\infty$, is positive.  Now,
$$
\lim_{h \to \infty} \left( \la - \al_1 \left( h - \sqrt{(r + h - \la + m)^2 + 4m \la} \right) \right) = \la + \al_1(r - \la + m),
$$
which one can show is positive.

Third, because $\al_2 < 0$ and $\be_1 > 1$, it is clear that $\ell(\al_2, \be_1) > 0$.

Fourth, when $h = 0$, $\ell(\al_2, \be_2) = \al_2 - \al_2 = 0$, and $\lim_{h \to \infty} \ell(\al_2, \be_2) = 0 - \left( {1 \over \la} \, (-\la) + 1 \right) \al_2 = 0$.  Next,
$$
{\partial \over \partial h} \ell(\al_2, \be_2) \propto \la - \al_2 \left( h - \sqrt{(r + h - \la + m)^2 + 4m \la} \right),
$$
which is negative when $h = 0$ and monotonically increases to a positive number as $h$ approaches $\infty$.  Thus, $\ell(\al_2, \be_2) < 0$ for all $h > 0$.  \qed

\lem{5.6} {If $c > rb$, then the following equation has a unique solution $x > 1:$
$$
\eqalign{
1 &= \left( {c - rb \over c(r + h)} \right)^{\al_1 - \al_2} \left({1 \over \al_1 - 1} \right)^{\al_1 - 1} \left( {1 \over 1 - \al_2} \right)^{1 - \al_2} \cr
& \quad \times \left[ h \left((\al_1 - 1) - {r \over \la} \, \al_1 \right) - {\ell(\al_1, \be_1) g(\be_2) \over \be_1 - \be_2} \, x^{\be_1 - 1} + {\ell(\al_1, \be_2) g(\be_1) \over \be_1 - \be_2} \, x^{\be_2 - 1} \right]^{\al_1 - 1} \cr
& \quad \times \left[ h \left((1 - \al_2) + {r \over \la} \, \al_2 \right) + {\ell(\al_2, \be_1) g(\be_2) \over \be_1 - \be_2} \, x^{\be_1 - 1} - {\ell(\al_2, \be_2) g(\be_1) \over \be_1 - \be_2} \, x^{\be_2 - 1} \right]^{1 - \al_2},
}
\eqno(5.15)
$$
in which $\al_1$, $\al_2$, and $\be_1$ are given in $(4.7)$ and $\be_2$ in $(5.2)$.}

\pf When $x = 1$, the right side of (5.15) equals $0$, and as $x \to \infty$, the right side approaches $\infty$.  Thus, if we show that the right side increases with $x$, then we are done.  Because $\ell(\al_1, \be_1) g(\be_2) < 0$, $\ell(\al_1, \be_2) g(\be_1) < 0$, $\be_1 > 1$, and $\be_2 < 0$, the factor on the second line of (5.15) increases with $x$.  Finally, 
$$
\eqalign{
&{d \over dx} \left(  \ell(\al_2, \be_1) g(\be_2) \, x^{\be_1 - 1} - \ell(\al_2, \be_2) g(\be_1) \, x^{\be_2 - 1} \right) \cr
& \quad \propto \ell(\al_2, \be_1) g(\be_2) (\be_1 - 1) \, x^{\be_1 - \be_2} + \ell(\al_2, \be_2) g(\be_1) (1 - \be_2),
}
$$
is positive for all $x > 1$ if it is positive when $x = 1$.  One can show that
$$
\ell(\al_2, \be_1) g(\be_2) (\be_1 - 1) + \ell(\al_2, \be_2) g(\be_1) (1 - \be_2) \propto \left( 1 - {h \over \la} \, \al_2 \right) (r \po - \la) - \al_2 m,
$$
and the right side is positive because $\al_2 < 0$ and $r \po - \la > 0$, in which $\po = p |_{h=0}$.  \qed

\thm{5.7} {If $h > {r \over r + m} \, \la$ and $rb < c < C_2$, then the maximum probability of reaching the bequest goal equals
$$
\phi(w) =
\cases{
{c \over r} \, {(\al_1 - 1)(1 - \al_2) \over \al_1 - \al_2} \left[ - \left( {y \over \yo} \right)^{\al_1} + \left( {y \over \yo} \right)^{\al_2} \right] \yo, &if $0 \le w < \wb = b - {\la \over h} {1 \over \yb}$, \cr \cr
1 - {c - rb \over r(r + h)} \, \left[{\be_1 - 1 \over \be_1 - \be_2} \, g(\be_2) \left( {y \over \yg} \right)^{\be_1} + {1 - \be_2 \over \be_1 - \be_2} \, g(\be_1) \left( {y \over \yg} \right)^{\be_2} \right] \yg
, &if $\wb \le w < b,$ \cr \cr
1 - \left( {c \over r} - b \right) {\yg \over \po} \left( {{c \over r} - w \over {c \over r} - b} \right)^{\po}, &if $b \le w \le \ws = {c \over r},$
}
\eqno(5.16)
$$
in which $\al_1$, $\al_2$, and $\be_1$ are given in $(4.7)$; $\be_2$ in $(5.2)$; $g$ in $(5.1)$; and $\po = p \big|_{h=0} = {\al_1 \over \al_1 - 1}$.  The parameter $\yb > 0$ is given by
$$
{\la \over h} \, {1 \over \yb} = b + {c \over r} \left[ {\al_1 (1 - \al_2) \over \al_1 - \al_2} \, \ybo^{\al_1 - 1} + {\al_2 (\al_1 - 1) \over \al_1 - \al_2} \, \ybo^{\al_2 - 1} - 1 \right] ,
\eqno(5.17)
$$
in which $\ybo \in (0, 1)$ is given by
$$
{c \over r} \, (1 - \al_2) \, \ybo^{\al_1 - 1} = {c - rb \over r(r+h)} \left[ h \left((1 - \al_2) + {r \over \la} \, \al_2 \right) + {\ell(\al_2, \be_1) g(\be_2) \over \be_1 - \be_2} \, \ybg^{\be_1 - 1} - {\ell(\al_2, \be_2) g(\be_1) \over \be_1 - \be_2} \, \ybg^{\be_2 - 1}  \right],
\eqno(5.18)
$$
and $\ybg > 1$ uniquely solves $(5.15)$.  The parameters $\yg$ and $\yo$ equal
$$
\yg = {\yb \over \ybg}  \hbox{ and } \yo = {\yb \over \ybo}.
$$
In the first expression of $(5.16)$, for a given $w \in [0, \wb)$, $y \in (\yb, \yo]$ uniquely solves $(4.11)$.  In the second expression of $(5.16)$, for a given $w \in [\wb, b)$, $y \in (\yg, \yb]$ uniquely solves $(5.7)$.
\vskip 0 pt
\indent When wealth equals $w \in (0, \ws]$, the optimal amount of instantaneous term life insurance equals
$$
D^*(w) = (b - w) \, {\bf 1}_{\{\wb \le w \le b\}},
\eqno(5.19)
$$
and the optimal amount invested in the risky asset equals
$$
\pi^*(w) = 
\cases{
{\mu - r \over \sig^2} \, {c \over r} \, {(\al_1 - 1)(1 - \al_2) \over \al_1 - \al_2} \left[ \al_1 \left( {y \over \yo} \right)^{\al_1 - 1} - \al_2 \left( {y \over \yo} \right)^{\al_2 - 1} \right], &if $0 \le w < \wb,$ \cr \cr
{\mu - r \over \sig^2} \, {c - rb \over r(r + h)} \left[ {\be_1 (\be_1 - 1) \over \be_1 - \be_2} g(\be_2) \left( {y \over \yg} \right)^{\be_1 - 1} + {\be_2 (1 - \be_2) \over \be_1 - \be_2} g(\be_1) \left( {y \over \yg} \right)^{\be_2 - 1} \right], &if $\wb \le w < b,$ \cr \cr
{\mu - r \over \sig^2} \, {{c \over r} - w \over \po - 1}, &if $b \le w \le \ws$.
}
\eqno(5.20)
$$}

\pf From Lemma 5.6, we know that there is a unique solution $\ybg > 1$ of (5.15). Next, we prove that $\ybo$ defined by (5.18) lies between $0$ and $1$.  It is easy to show that $(1 - \al_2) + {r \over \la} \, \al_2$ is positive; thus, the right side of (5.18) is positive, so $\ybo > 0$.  One can use (5.15) and (5.18) to show that $\ybo = 1$ if and only if $c = C_2$, the unique solution of (5.3).  Thus, to prove that $\ybo < 1$, it is enough to prove that $\ybo$ in (5.18) increases with $c$.  To that end, compute ${\partial \ybo \over \partial c}$ via (5.18), compute ${\partial \ybg \over \partial c}$ via (5.15), and solve for ${\partial \ybo \over \partial c}$ to obtain
$$
{\partial \ybo \over \partial c} \propto {N_1 D_2 + N_2 D_1 \over (\al_1 - 1) N_1 D_2 - (1 - \al_2) N_2 D_1},
\eqno(5.21)
$$ 
in which 
$$
N_1 = (\be_1 - 1) \ell(\al_1, \be_1) g(\be_2) \, \ybg^{\be_1 - 1} + (1 - \be_2) \ell(\al_1, \be_2) g(\be_1) \, \ybg^{\be_2 - 1},
$$
$$
N_2 = (\be_1 - 1) \ell(\al_2, \be_1) g(\be_2) \, \ybg^{\be_1 - 1} + (1 - \be_2) \ell(\al_2, \be_2) g(\be_1) \, \ybg^{\be_2 - 1},
$$
$$
D_1 = h (\be_1 - \be_2) \left( (\al_1 - 1) - {r \over \la} \, \al_1 \right) - \ell(\al_1, \be_1) g(\be_2) \, \ybg^{\be_1 - 1} + \ell(\al_1, \be_2) g(\be_1) \, \ybg^{\be_2 - 1},
$$
and
$$
D_2 = h (\be_1 - \be_2) \left( (1 - \al_2) + {r \over \la} \, \al_2 \right) + \ell(\al_2, \be_1) g(\be_2) \, \ybg^{\be_1 - 1} - \ell(\al_2, \be_2) g(\be_1) \, \ybg^{\be_2 - 1}.
$$
The numerator on the right side of (5.21) simplifies to the following.
$$
\eqalign{
N_1 D_2 + N_2 D_1 &\propto h (\be_1 - 1) \left( (\be_1 - 1) - {r + h \over \la} \, \be_1 \right) g(\be_2) \, \ybg^{\be_1 - 1} \cr
& \quad - h(1 - \be_2) \left( (1 - \be_2) + {r + h \over \la} \, \be_2 \right) g(\be_1) \, \ybg^{\be_2 - 1} \cr
& \quad - (\be_1 - \be_2) g(\be_1) g(\be_2) \, \ybg^{\be_1 - 1} \, \ybg^{\be_2 - 1}.
}
$$
It is straightforward to show that $\left( (\be_1 - 1) - {r + h \over \la} \, \be_1 \right) < 0$ and $\left( (1 - \be_2) + {r + h \over \la} \, \be_2 \right) > 0$; thus, $N_1 D_2 + N_2 D_1 < 0$, from which it follows
$$
{\partial \ybo \over \partial c} \propto  (1 - \al_2) N_2 D_1 - (\al_1 - 1) N_1 D_2.
\eqno(5.23)
$$
If we show that $N_1 < 0$, $N_2 > 0$, $D_1 > 0$, and $D_2 > 0$, then we will have shown that $\ybo$ increases with $c$.  We prove these four inequalities in turn.

First, $N_1 < 0$ follows directly from $\ell(\al_1, \be_1) < 0$ and $\ell(\al_1, \be_2) < 0$, which we proved in Lemma 5.5, and from $g(\be_1) > 0$ and $g(\be_2) > 0$, which we proved in Lemma 5.1.

Second, $N_2 > 0$ is equivalent to
$$
(\be_1 - 1) \ell(\al_2, \be_1) g(\be_2) \, x^{\be_1 - \be_2} + (1 - \be_2) \ell(\al_2, \be_2) g(\be_1) > 0,
$$
for all $x > 1$, which is equivalent to
$$
(\be_1 - 1) \ell(\al_2, \be_1) g(\be_2) + (1 - \be_2) \ell(\al_2, \be_2) g(\be_1) \ge 0,
\eqno(5.22)
$$
because $(\be_1 - 1) \ell(\al_2, \be_1) g(\be_2) > 0$. Inequality (5.22) is straightforward, but tedious, to demonstrate.

Third, $D_1$ increases with $\ybg > 1$; thus, we only need to show that $D_1 \ge 0$ when $\ybg = 1$.  Because $D_1 \big|_{\ybg = 1} = 0$, it follows that $D_1 > 0$ for all $\ybg > 1$.

Fourth, $D_2 > 0$ follows directly from  $\left( (1 - \al_2) + {r \over \la} \, \al_2 \right) > 0$, from $\ell(\al_2, \be_1) > 0$ and $\ell(\al_2, \be_2) < 0$, which we proved in Lemma 5.5, and from $g(\be_1) > 0$ and $g(\be_2) > 0$, which we proved in Lemma 5.1.  Thus, we have proved that $\ybo$ increases with $c$.

Next, we prove that $\yb > 0$.  Because $\ybo$ increases with $c$, the right side of (5.17) increases with $c$, so it is enough to show that the right side of (5.17) is positive as $c \to rb+$.  From (5.15), we deduce that
$$
\lim_{c \to rb+} (c - rb) \ybg^{\be_1 - 1} = \left( - {\al_1 - 1 \over \ell(\al_1, \be_1)} \right)^{{\al_1 - 1 \over \al_1 - \al_2}} \left( {1 - \al_2 \over \ell(\al_2, \be_1)} \right)^{{1 - \al_2 \over \al_1 - \al_2}} \, {\be_1 - \be_2 \over g(\be_2)} \, rb(r+h),
$$
which implies that
$$
\lim_{c \to rb+} \ybo = \left( - {\al_1 - 1 \over 1 - \al_2} \, {\ell(\al_2, \be_1) \over \ell(\al_1, \be_1)} \right)^{{1 \over \al_1 - \al_2}}.
\eqno(5.23)
$$
Thus, the limit of the right side of (5.17), as $c \to rb+$, equals
$$
\eqalign{
&b \al_1 (1 - \al_2) \left( - {\al_1 - 1 \over 1 - \al_2} \, {\ell(\al_2, \be_1) \over \ell(\al_1, \be_1)} \right)^{{al_1 - 1 \over \al_1 - \al_2}} + b \al_2 (\al_1 - 1) \left( - {\al_1 - 1 \over 1 - \al_2} \, {\ell(\al_2, \be_1) \over \ell(\al_1, \be_1)} \right)^{-{1 - \al_2 \over \al_1 - \al_2}} \cr
&\; \propto \al_1 \ell(\al_2, \be_1) - \al_2 \ell(\al_1, \be_1) = \be_1(\al_1 - \al_2) > 0.}
$$
From $\yb > 0$, it follows that $\wb = b - {\la \over h} \, {1 \over \yb} < b$.

Next, we prove that $\wb > 0$.  From the expression in (5.17), we obtain the same expression for $\wb$, as a function of $\ybo$, as the one in (4.14).  The expression in square brackets in (4.14) decreases with $c$ because $\ybo$ increases with $c$.  Thus, to show that $\wb > 0$, it is enough to show that
$$
\lim_{c \to rb+} 1 - {\al_1 (1 - \al_2) \over \al_1 - \al_2} \, \ybo^{\al_1 - 1} - {\al_2 (\al_1 - 1) \over \al_1 - \al_2} \, \ybo^{\al_2 - 1} \ge 0,
$$
which, from (5.23), is equivalent to
$$
\left( - {\al_1 - 1 \over \ell(\al_1, \be_1)} \right)^{\al_1 - 1} \left( {1 - \al_2 \over \ell(\al_2, \be_1)} \right)^{1 - \al_2} \le 1. 
\eqno(5.24)
$$
The term $- \ell(\al_1, \be_1)$ is positive and increases with $h$. Thus, to show that the first factor on the left side of (5.24) is less than $1$ for all $h > {r \over r + m} \, \la$, it is enough to show that the first factor is less than or equal to $1$ for $h = {r  \over r +m} \, \la$.
$$
- {\al_1 - 1 \over \ell(\al_1, \be_1)} \bigg|_{h = {r \over r + m} \, \la} \le 1
$$
is equivalent to
$$
{hp \over \la} \bigg|_{h = {r \over r + m} \, \la} \cdot \al_1 \ge 1,
$$
which is true because the left side reduces to $\al_1$, and we know that $\al_1 > 1$.  Also, the second factor on the left side of (5.24) is less than $1$ for all $h \ge 0$.  Thus, we have proved that $\wb > 0$.

The rest of the proof is similar to the proofs of Theorems 4.3 and 5.2, so we omit those details.  \qed

\rem{5.2} {As noted in Remark 4.3 for the case studied in Section 4.2, when $0 < w < \wb$, the optimal amount invested in the risky asset is independent of both $b$ and $h$, a surprising myopic result. \qed}

Observe that the solution in Theorem 5.7 is continuous as the bequest goal approaches 0, as in Corollary 5.3 for the case in Section 5.1.

\cor{5.8} {The solution given in Theorem $5.7$ is continuous as $b \to 0+$.  In particular, for $0 \le w \le {c \over r}$,
$$
\lim_{b \to 0+} \phi(w) = 1 - \left( 1 - {rw \over c} \right)^{\po},
$$
and
$$
\lim_{b \to 0+} \pi^*(w) = {\mu - r \over \sig^2} \, {{c \over r} - w \over \po - 1}.
$$}

In the next corollary, we observe that the optimal amount invested in the risky asset decreases with wealth if $w \ge \wb$, as shown in Corollary 5.4 for the case in Section 5.1; there, $\wb = 0$.  We omit the proof because it is identical to the proof of Corollary 5.4.  Also, Remark 5.1 applies in this case.

\cor{5.9} {If $h > {r \over r + m} \, \la$ and $rb < c < C_2$, then the optimal amount invested in the risky asset decreases as wealth increases for $\wb \le w \le \ws$.  \qed}

%\rem{5.3} {As observed in Remark 5.1, $\pi^*$ for wealth between $b$ and ${c \over r}$ is {\it identical} to the amount invested in the risky asset when minimizing the probability of lifetime ruin (Young, 2004).  Also, as in Young (2004) and as in the cases considered in Sections 3, 4, and 5.1, wealth never reaches the safe level.\qed}

\sect{6. Properties of $\phi$, $D^*$, and $\pi^*$}

In this section, we prove general properties of the solution obtained in Sections 3 through 5.  As the premium rate for life insurance increases, we expect the maximum probability of reaching the bequest goal to decrease because it becomes more difficult for the individual to reach her bequest goal.  We demonstrate this in the following proposition and find the limits of $\phi$ as $h \to 0+$ and $h \to \infty$.

\prop{6.1} {The maximum probability of reaching the bequest goal $($weakly$)$ decreases with $h$.  Furthermore,
$$
\lim_{h \to 0+} \phi(w) = 1 - \left( 1 - {rw \over c} \right)^{\po}, \quad 0 \le w \le {c \over r},
\eqno(6.1)
$$
in which
$$
\po = p \big|_{h=0} = {1 \over 2r} \left[ (r + \la + m) + \sqrt{(r + \la + m)^2 - 4r \la} \right] > 1,
\eqno(6.2)
$$
and
$$
\lim_{h \to \infty} \phi(w) = \phio(w), \quad 0 \le w \le \max \left( {c \over r}, \, b \right),
\eqno(6.3)
$$
in which $\phio$ is the maximum probability of reaching the bequest goal when life insurance is not available $($Bayraktar and Young, 2015$)$.}

\pf  In Sections 3 through 5, we showed that $\phi$ in (2.2) is a classical solution of its HJB equation (2.9).  Define $F$ by
$$
F(w, f, f_w, f_{ww}; h) = \la f - (rw - c) f_w - \max_\pi \left[ (\mu - r) \pi \, f_w + {1 \over 2} \sigma^2 \pi^2 f_{ww} \right] - (\la - h(b - w)_+ f_w)_+.
$$
Note that $F$ increases with $f$ and decreases with $f_{ww}$; thus, $F$ satisfies the monotonicity condition (0.1) in Crandall et al.\ (1992).  Suppose $h^1 < h^2$, and let $\phi^{(i)}$ denote the maximum probability of reaching the bequest goal when $h = h^i$ for $i = 1, 2$, with corresponding safe level $\ws^{(i)}$.  Note that $\ws^{(1)} \le \ws^{(2)}$.  We have $F \left(w, \phi^{(i)}, \phi^{(i)}_w, \phi^{(i)}_{ww}; h^i \right) = 0$ for $i = 1, 2$, and
$$
F \left(w, \phi^{(2)}, \phi^{(2)}_w, \phi^{(2)}_{ww}; h^1 \right) = \left(\la - h^2(b - w)_+ \phi^{(2)}_w \right)_+ - \left(\la - h^1(b - w)_+ \phi^{(2)}_w \right)_+ \le 0,
$$
because $\phi$ increases with $w$.  Thus, $\phi^{(2)}$ is a viscosity subsolution of $F \left(w, \phi, \phi_w, \phi_{ww}; h^1 \right) = 0$.  Because $\phi^{(1)}$ is a classical solution of this equation, because $\phi^{(2)}(0) = \phi^{(1)}(0)$, and because $\phi^{(2)} \left(\ws^{(1)} \right) \le 1 = \phi^{(1)} \left(\ws^{(1)} \right)$, it follows from Crandall et al.\ (1992, Theorem 3.3) that $\phi^{(2)} \le \phi^{(1)}$ on $\left[0, \ws^{(1)} \right]$.

Furthermore, from the stability of viscosity solutions, we can find the limit of $\phi$ as $h \to 0+$ or $h \to \infty$ by taking the corresponding limit of the HJB equation.  To that end, note that as $h \to 0+$, (2.9) becomes
$$
\left\{
\eqalign{
&\la (\Phi - 1) = (rw - c) \Phi_w + \max_\pi \left[ (\mu - r) \pi \, \Phi_w + {1 \over 2} \sigma^2 \pi^2 \Phi_{ww} \right], \cr
&\Phi(0) = 0, \quad \Phi(c/r) = 1,
}
\right.
\eqno(6.4)
$$ 
because $\lim_{h \to 0+} h \phi_w(w) = 0$ for all $w \in (0, b)$, and the solution of this BVP is given in (6.1).  Also, note that as $h \to \infty$, (2.9) becomes
$$
\left\{
\eqalign{
&\la (\Phi - {\bf 1}_{\{w \ge b\}}) = (rw - c) \Phi_w + \max_\pi \left[ (\mu - r) \pi \, \Phi_w + {1 \over 2} \sigma^2 \pi^2 \Phi_{ww} \right] , \cr
&\Phi(0) = 0, \quad \Phi(\max(c/r, b)) = 1.
}
\right.
\eqno(6.5)
$$
The BVP in (6.5) is the one solved by $\phio$, as computed in Bayraktar and Young (2015).  \qed

\rem{6.1} {As $h$ approaches $0$, the maximum probability of reaching the bequest goal approaches $1$ minus the minimum probability of lifetime ruin (Young, 2004).  We expect this result because, as $h$ approaches $0$, covering the bequest goal becomes costless, so the problem reduces to one of avoiding ruin. \qed}

\rem{6.2} {From the solution to the optimization problem given in Theorems 4.2 and 5.2, it is optimal to buy insurance for all levels of wealth less than $b$ if either of the following holds:  \hfill \break
(a) $h \le {r \over r + m} \, \la$ and $c \ge C_1$; or \hfill \break
(b) $h > {r \over r + m} \, \la$ and $c \ge C_2$.  \hfill \break
Thus, it is optimal to buy life insurance for all levels of wealth less than $b$ if the rate of consumption is large enough, in which large enough depends on $h$.  \qed}

As the premium rate becomes arbitrarily small, covering the bequest goal with life insurance becomes costless, and we expect the buying level to become arbitrarily small.  By contrast, as the premium rate becomes arbitrarily large, the individual will not purchase life insurance, that is, we expect the buying level to approach $b$.  These limits are easy to prove, so we present the next proposition without proof.

\prop{6.2} {The buying level for insurance $\wb$ obeys the following limits:
$$
\lim_{h \to 0+} \wb = 0,
$$
and
$$
\lim_{h \to \infty} \wb = b.
\eqno{\square}
$$}

As the premium rate for life insurance increases, we expect the amount invested in the risky asset to increase because the individual has to take on more financial risk to reach her bequest goal.  We demonstrate this in the following proposition and find the limits of $\pi^*$ as $h \to 0+$ and $h \to \infty$.

\prop{6.3} {The optimal amount invested in the risky asset $($weakly$)$ increases as $h$ increases.  Furthermore,
$$
\lim_{h \to 0+} \pi^*(w) = \pi^{min}(w), \quad 0 \le w \le {c \over r},
\eqno(6.6)
$$
in which $\pi^{min}$ is the optimal amount invested in the risky asset when minimizing the probability of lifetime ruin,
specifically,
$$
\pi^{min}(w) = {\mu - r \over \sig^2} \, {{c \over r} - w \over \po - 1},
$$
and
$$
\lim_{h \to \infty} \pi^*(w) = \pio(w), \quad 0 \le w \le \max \left( {c \over r}, \, b \right),
\eqno(6.7)
$$
in which $\pio$ is the optimal amount invested in the risky asset when life insurance is not available $($Bayraktar and Young, 2015$)$.}

\pf  From the solution in Sections 3 through 5, we know that $\pi^*(w)$ is independent of $h$ for $0 \le w < \wb$ (although $\wb$ itself depends on $h$) and for $b \le w \le {c \over r}$; the latter applies to the solution in Section 5 only.  For $\wb < w < \min(\ws, b)$, we will find a differential equation for $\pi^*$ and use comparison to show that $\pi^*$ increases with $h$.  To that end, from the solution in Sections 3 through 5, we know that $\phi$ solves the following differential equation for $\wb < w < \min(\ws, b)$:
$$
\la (\phi - 1) = ((r+h)w - (c+hb)) \phi_w - m \, {\phi_w^2 \over \phi_{ww}},
$$
or equivalently,
$$
\la (\phi - 1) = ((r+h)w - (c+hb)) \phi_w + {\mu - r \over 2} \, \phi_w \, \pi^*,
$$
because $\pi^* = - {\mu - r \over \sig^2} \, {\phi_w \over \phi_{ww}}$.  Differentiate this expression with respect to $w$ and rewrite the result to obtain a differential equation for $\pi^*$:
$$
{\mu - r \over 2} \, \pi^*_w + {\mu - r \over \sig^2} \, {(c + hb) - (r + h)w \over \pi^*} + (r + h - \la - m) = 0.
\eqno(6.8)
$$

%Let $0 < h^1 < h^2$ and write $\pi^{(i)}(w)$ for $\pi^*(w; h^i)$, $i = 1, 2$. We wish to show that $\pi^{(1)}(w) \le \pi^{(2)}(w)$ for 

Differentiate (6.8) with respect to $h$ to obtain a differential equation for $\pi^*_h$:
$$
{\mu - r \over 2} \, (\pi^*_h)_w - {\mu - r \over \sig^2} \, {(c + hb) - (r + h)w \over (\pi^*)^2} \, \pi^*_h + {\mu - r \over \sig^2} \, {b - w \over \pi^*} + 1 = 0.
\eqno(6.9)
$$
Redefine the independent variable so that $\min(\ws, b)$ becomes the new origin: $\tw := \min(\ws, b) - w$, and $\tpi(\tw) := \pi^*(\min(\ws, b) - \tw)$.  Then, $(c + hb) - (r + h)w = (r + h) \tw + (c - rb)_+$, $b - w = \tw + (b - \ws)_+ $, $\pi^*_h = \tpi_h$, $(\pi^*_h)_w = - (\tpi_h)_{\tw}$, and (6.9) becomes
$$
{\mu - r \over 2} \, (\tpi_h)_{\tw} + {\mu - r \over \sig^2} \, {(r + h) \tw + (c - rb)_+ \over (\tpi)^2} \; \tpi_h - {\mu - r \over \sig^2} \, {\tw + (b - \ws)_+ \over \tpi} - 1 = 0.
$$
Define $G$ by
$$
G(\tw, f, f_{\tw}) = {\mu - r \over 2} \, f_{\tw} + {\mu - r \over \sig^2} \, {(r + h) \tw + (c - rb)_+ \over (\tpi)^2} \; f - {\mu - r \over \sig^2} \, {\tw + (b - \ws)_+ \over \tpi} - 1.
$$
Note that $G$ increases with $f$; thus, $G$ satisfies the monotonicity condition (0.1) in Crandall et al. (1992). Then, $G(\tw, \tpi_h, (\tpi_h)_{\tw}) = 0$, and
$$
G(\tw, 0, 0) = - {\mu - r \over \sig^2} \, {\tw + (b - \ws)_+ \over \tpi} - 1 < 0.
$$
After a great deal of algebra, one can show that $\tpi_h(0+) = \pi^*_h(\min(\ws, b)-) \ge 0$; then, from Crandall et al.\ (1992, Theorem 3.3), we conclude that $\pi^*_h(w) \ge 0$ for $\wb < w < \min(\ws, b)$. 

One can show the limits in (6.6) and (6.7) on a case-by-case basis by using the solution given in Sections 3 through 5.  In the interest of space, we do not include those calculations here.  \qed

Proposition 6.3 tells us that $\pi^* \ge \pi^{min}$ because the problem of minimizing the probability of lifetime ruin is equivalent to the problem we obtain as $h \to 0+$.  Furthermore, the expressions for $\pi^*$ in (5.9) and (5.20) show us that $\pi^*(w) = \pi^{min}(w)$ for all $b \le w \le {c \over r}$.

Bayraktar and Young (2015) compute the optimal investment strategy to maximize the probability of reaching the bequest goal when life insurance is not available.  Proposition 6.3 tells us that $\pi^* \le \pio$.  By comparing the results in Bayraktar and Young (2015) with the solution here when it is optimal not to purchase life insurance, we have the following proposition.

\prop{6.4} {Let $\pi^{min}$ and $\pio$ denote the optimal investment strategies to minimize the probability of lifetime ruin and to maximize the probability of reaching the bequest goal when life insurance is not available, respectively.  Then,
$$
\pi^{min}(w) \le \pi^*(w) = \pio(w), \quad 0 \le w < \wb,
\eqno(6.10)
$$
$$
\pi^{min}(w) \le \pi^*(w) \le \pio(w), \quad \wb \le w < b,
\eqno(6.11)
$$
and
$$
\pi^{min}(w) = \pi^*(w) = \pio(w), \quad b \le w \le {c \over r},
\eqno(6.12)
$$
with the understanding that $\pi^*(w) = 0$ if $w \ge \ws$.  \qed}

\rem{6.3} {The investment strategy when it is optimal not to buy life insurance ($0 \le w < \wb$ or $b \le w \le {c \over r}$) is myopic because the individual is seemingly indifferent to the presence or lack of life insurance.  On the other hand, when it is optimal to buy life insurance ($\wb \le w <  b$), the presence of life insurance leads the individual to invest less in the risky asset than when life insurance is not available.  To reach the bequest goal, the individual does not have to take on as much risk when life insurance is available.  \hfill \break
\indent  Life insurance allows the individual to achieve her bequest goal without the necessity of wealth reaching the bequest target itself.  If no life insurance is available, the {\it only} way the individual will reach her bequest goal is if wealth itself reaches that bequest goal $b$.  \hfill \break
\indent We find it interesting that the optimal investment strategy when wealth is greater than the bequest goal $b$ is {\it identical} to the corresponding one for minimizing the probability of lifetime ruin, which is independent of the ruin level.  Once wealth is greater than the bequest goal $b$, our individual invests as if she were minimizing the probability of lifetime ruin with ruin level $b$, or any ruin level, for that matter (Bayraktar and Young, 2007). \qed}

We end this section with two numerical examples.  In the following example, we demonstrate how the optimal strategies change as $c$ increases.

\ex{6.1} {Consider the following parameters values: $r = 0.03$, $\mu = 0.06$, $\sig = 0.20$, $\la = 0.04$, $h = 0.05$, and $b = 1.0$.  Thus, $C_1 = 0.0736$, as defined by equation (4.1), and $C_2 = 0.0629$, as defined by equation (5.3); note that $h >  {r \over r + m} \, \la$.  We have the following table that displays how the optimal strategies $\wb$ and $\pi^*$ vary as $c$ increases from $0$ to $C_2$.  We set $\pi^*(w) = 0$ when $w \ge \ws$.
$$\vbox{\settabs 8 \columns
\+ $c$ & $\wb$ & $\ws$ & $\pi^*(0.1)$ & $\pi^*(0.3)$ & $\pi^*(0.5)$ & $\pi^*(0.7)$ & $\pi^*(0.9)$ \cr \smallskip
\hrule \smallskip
\+ $0$ & $0.375$ & $0.625$ & $0.212$ & $0.637$ & $0.397$ & 0 & 0 \cr
\+ $0.0005$ & $0.381$ & $0.631$ & $0.207$ & $0.622$ & $0.417$ & 0 & 0 \cr
\+ $0.005$ & $0.403$ & $0.688$ & $0.428$ & $0.724$ & $0.560$ & 0 & 0 \cr
\+ $0.01$ & $0.397$ & $0.750$ & $0.748$ & $0.983$ & $0.794$ & $0.159$ & 0 \cr
\+ $0.02$ & $0.354$ & $0.875$ & $1.407$ & $1.597$ & $1.191$ & $0.556$ & 0 \cr
\+ $0.03$ & $0.295$ & $1.000$ & $2.072$ & $2.223$ & $1.588$ & $0.953$ & $0.318$ \cr
\+ $0.04$ & $0.215$ & $1.333$ & $2.693$ & $2.575$ & $1.932$ & $1.284$ & $0.615$ \cr
\+ $0.05$ & $0.124$ & $1.667$ & $3.359$ & $2.893$ & $2.239$ & $1.573$ & $0.874$ \cr
\+ $0.06$ & $0.028$ & $2.000$ & $3.851$ & $3.194$ & $2.528$ & $1.846$ & $1.122$ \cr
\+ $0.0629$ & $0$ & $2.097$ & $3.937$ & $3.278$ & $2.609$ & $1.923$ & $1.193$ \cr \smallskip \hrule}
$$
Note that, in this example, even though $h > {r \over r + m} \, \la = 0.029\overline{09}$, $\wb$ first increases from $0.375$ to $0.403$ and then decreases to $0$ as $c$ increases from $0$ to $C_2$.

Recall that the dollar amounts invested in the risky asset are relative to a bequest goal of $b = 1$, so one can think of the dollar amounts as proportions of the bequest goal, as discussed in Remark 2.5.   We see that $\pi^*$ is not monotone in $c$ and is not monotone in $w$ for $0 < w < \wb$.  However, from the expressions in (4.4) and (4.13) and from Corollaries 5.4, and 5.9, we know that $\pi^*$ decreases with $w$ for $w > \wb$.  Also note that, in this example, $\pi^*$ eventually increases with $c$ because the individual needs to invest more in the risky asset to cover her additional consumption.  \qed}

\ex{6.2} {Continue with the parameter values of the Example 6.1, except that we will examine how the optimal strategies vary with the price of insurance $h$; recall that $\la = 0.04$ and ${r \over r + m} \, \la = 0.029\overline{09}$.    We have the following table for $c = 0.02 < rb = 0.03$:
$$\vbox{\settabs 8 \columns
\+ $h$ & $\wb$ & $\ws$ & $\pi^*(0.1)$ & $\pi^*(0.3)$ & $\pi^*(0.5)$ & $\pi^*(0.7)$ & $\pi^*(0.9)$ \cr \smallskip
\hrule \smallskip
\+ $0$ & $0$ & $0.667$ & $0.400$ & $0.259$ & $0.118$ & 0 & 0 \cr
\+ $0.01$ & $0$ & $0.750$ & $0.707$ & $0.490$ & $0.272$ & $0.0544$ & 0 \cr
\+ $0.02$ & $0$ & $0.800$ & $1.078$ & $0.770$ & $0.462$ & $0.154$ & 0 \cr
\+ $0.03$ & $0.133$ & $0.833$ & $1.407$ & $1.092$ & $0.683$ & $0.273$ & 0 \cr
\+ $0.04$ & $0.259$ & $0.857$ & $1.407$ & $1.447$ & $0.927$ & $0.408$ & 0 \cr
\+ $0.05$ & $0.357$ & $0.875$ & $1.407$ & $1.600$ & $1.191$ & $0.556$ & 0 \cr
\+ $0.10$ & $0.609$ & $0.923$ & $1.407$ & $1.600$ & $1.833$ & $1.402$ & $0.145$ \cr
\+ $0.20$ & $0.782$ & $0.957$ & $1.407$ & $1.600$ & $1.833$ & $2.106$ & $0.724$ \cr
\+ $0.50$ & $0.907$ & $0.981$ & $1.407$ & $1.600$ & $1.833$ & $2.106$ & $2.406$ \cr
\+ $\infty$ & $b = 1$ & $1.000$ & $1.407$ & $1.600$ & $1.833$ & $2.106$ & $2.406$ \cr \smallskip \hrule}
$$
Note that $\pi^*$ (weakly) increases with $h$, as expected from Proposition 6.3.  Also, $\pi^*$ is independent of $h$ for $0 \le w < \wb$, as expected from (6.10) in Proposition 6.4.  \qed}

%We have the following table for $c = 0.04 > rb = 0.03$:
%$$\vbox{\settabs 8 \columns
%\+ $h$ & $\wb$ & $\pi^*(0.1)$ & $\pi^*(0.3)$ & $\pi^*(0.5)$ & $\pi^*(0.7)$ & $\pi^*(0.9)$ & $\pi^*(1.1)$ \cr \smallskip
%\hrule \smallskip
%\+ $0$ & $0$ & $0.871$ & $0.730$ & $0.589$ & $0.447$ & $0.306$ & $0.165$ \cr
%\+ $0.01$ & $0$ & $1.249$ & $1.031$ & $0.811$ & $0.590$ & $0.362$ & $0.165$ \cr
%\+ $0.02$ & $0$ & $1.686$ & $1.376$ & $1.064$ & $0.749$ & $0.422$ & $0.165$ \cr
%\+ $0.03$ & $0$ & $2.167$ & $1.754$ & $1.339$ & $0.920$ & $0.485$ & $0.165$ \cr
%\+ $0.04$ & $0.079$ & $2.680$ & $2.156$ & $1.630$ & $1.099$ & $0.550$ & $0.165$ \cr
%\+ $0.05$ & $0.215$ & $2.693$ & $2.575$ & $1.932$ & $1.284$ & $0.615$ & $0.165$ \cr
%\+ $0.10$ & $0.551$ & $2.682$ & $2.719$ & $2.764$ & $2.240$ & $0.945$ & $0.165$ \cr
%\+ $0.20$ & $0.759$ & $2.675$ & $2.700$ & $2.727$ & $2.762$ & $1.598$ & $0.165$ \cr
%\+ $0.50$ & $0.900$ & $2.671$ & $2.684$ & $2.704$ & $2.729$ & $3.568$ & $0.165$ \cr
%\+ $\infty$ & $b = 1$ & $2.668$ & $2.675$ & $2.689$ & $2.707$ & $2.729$ & $0.165$ \cr \smallskip \hrule}
%$$

%From Proposition 6.4, we know that properties of $\pio$ also hold for $\pi^*$ when $0 \le w < \wb$ and $b \le w \le {c \over r}$.  See Bayraktar and Young (2015) for properties of the optimal investment strategy $\pio$; we do not reproduce that work here.

\sect{7. Summary}

We determined the optimal strategies for purchasing instantaneous term life insurance and for investing in a risky asset in order to maximize the probability of reaching a specific bequest goal $b$.  We proved the following properties of these optimal strategies and the corresponding maximum probability.

\smallskip

\item{$\bullet$} The premium rate for life insurance $h$ acts as a parameter to connect two seemingly unrelated problems.  First, as $h \to 0+$, the problem becomes equivalent to minimizing the probability of lifetime ruin.  Second, as $h \to \infty$, the problem becomes equivalent to maximizing the probability of reaching the bequest goal without life insurance in the market.  See Propositions 6.1 through 6.4 for focused results about this connection.

\smallskip

\item{$\bullet$} As in the problem of minimizing the probability of lifetime ruin (Young, 2004), optimally controlled wealth never reaches the safe level.

\smallskip

\item{$\bullet$} As $h$ increases, the maximum probability of reaching the bequest goal (weakly) decreases, and the optimal amount invested in the risky asset (weakly) increases because one has to take on more risk in the financial market to reach the bequest goal if one does not buy life insurance; see Propositions 6.1 and 6.3.

\smallskip

\item{$\bullet$} It is optimal to buy life insurance for all levels of wealth (less than $b$) if the rate of consumption is large enough, in which large enough depends on the premium rate; see Theorems 4.2 and 5.5, as well as Remark 6.2.  This result is surprising because if wealth is close to zero and one buys insurance, then the probability of ruin is greater than if one does not buy insurance.  However, the goal is not simply not to ruin; the goal is to reach the bequest $b$, which can only be achieved by buying sufficient life insurance if wealth is small.

\smallskip

\item{$\bullet$} It is optimal to buy life insurance only when wealth lies between a positive buying level $\wb > 0$ and $b$ if either of the following conditions holds:
\item\item{(a)} $h \le {r \over r + m} \, \la$ and $0 \le c < C_1$, or
\item\item{(b)} $h > {r \over r + m} \, \la$ and $0 \le c < C_2$.

\hangindent 20 pt Thus, it is optimal not to buy life insurance if one is poor and if the rate of consumption is small enough, in which `small enough' depends on the premium rate.

\smallskip

\item{$\bullet$} When it is optimal to buy life insurance, the optimal amount invested in the risky asset decreases with wealth (sometimes linearly), which is the case when minimizing the probability of lifetime ruin.  This result makes sense because if one is purchasing life insurance, then the bequest goal is covered, and the remaining problem is to avoid ruin.  In fact, for wealth greater than the bequest goal $b$, the optimal amount to invest in the risky asset is {\it identical} to the corresponding amount when minimizing the probability of lifetime ruin; see Proposition 6.4 and Remark 6.3.

\smallskip

\item{$\bullet$} When it is optimal not to buy life insurance, the optimal amount invested in the risky asset is {\it independent} of both the bequest goal and the price of life insurance and is {\it identical} to the corresponding amount when maximizing the probability of reaching the bequest goal without life insurance in the market; see Proposition 6.4.  We were surprised by this myopic investment behavior.

\bigskip

\centerline{\bf Acknowledgments} \medskip  We thank two anonymous referees and an associated editor for their helpful comments.  Research of the first author is supported in part by the National Science Foundation under grant DMS-0955463 and by the Susan M. Smith Professorship of Actuarial Mathematics. Research of the third author is supported in part by the Cecil J. and Ethel M. Nesbitt Professorship of Actuarial Mathematics.

\sect{References}

%\noindent \hangindent 20 pt Bayraktar, Erhan, Xueying Hu, and Virginia R. Young (2011), Minimizing the probability of lifetime ruin under stochastic volatility, {\it Insurance: Mathematics and Economics}, 49 (2): 194Ð206.

\noindent \hangindent 20 pt Bayraktar, Erhan, S. David Promislow, and Virginia R. Young (2014), Purchasing life insurance to reach a bequest goal, {\it Insurance: Mathematics and Economics}, 58: 204-216.

\smallskip \noindent \hangindent 20 pt Bayraktar, Erhan, S. David Promislow, and Virginia R. Young (2015), Purchasing life insurance to reach a bequest goal: time-dependent case, to appear in {\it North American Actuarial Journal}.

\smallskip \noindent \hangindent 20 pt Bayraktar, Erhan and Virginia R. Young (2007), Correspondence between lifetime minimum wealth and utility of consumption, {\it Finance and Stochastics}, 11 (2): 213- 236.

\smallskip \noindent \hangindent 20 pt Bayraktar, Erhan and Virginia R. Young (2009), Minimizing the lifetime shortfall or shortfall at death, {\it Insurance: Mathematics and Economics}, 44 (3): 447-458.

\smallskip \noi \hangindent 20 pt Bayraktar, Erhan and Virginia R. Young (2013), Life insurance purchasing to maximize utility of household consumption, {\it North American Actuarial Journal}, 17 (2): 114-135.

\smallskip \noi \hangindent 20 pt Bayraktar, Erhan and Virginia R. Young (2015), Optimally investing to reach a bequest goal, working paper, University of Michigan.

\smallskip \noindent \hangindent 20 pt Browne, Sid (1997), Survival and growth with a liability: optimal portfolio strategies in continuous time, {\it Mathematics of Operations Research}, 22 (2): 468-493.

\smallskip \noindent \hangindent 20 pt Browne, Sid (1999a), Beating a moving target: optimal portfolio strategies for outperforming a stochastic benchmark, {\it Finance and Stochastics}, 3 (3): 275-294.

\smallskip \noindent \hangindent 20 pt Browne, Sid (1999b), Reaching goals by a deadline: digital options and continuous-time active portfolio management, {\it Advances in Applied Probability}, 31 (2): 551-577.

%\smallskip \noindent \hangindent 20 pt Bowers, Newton L., Hans U. Gerber, James C. Hickman, Donald A. Jones, and Cecil J. Nesbitt (1997), {\it Actuarial Mathematics}, second edition, Schaumburg, IL: Society of Actuaries.

%\smallskip \noindent \hangindent 20 pt Campbell, Ritchie A. (1980), The demand for life insurance: an application of the economics of uncertainty, {\it Journal of Finance}, 35 (5), 1155-1172.

\smallskip \noindent \hangindent 20 pt Crandall, Michael G., Hitoshi Ishii, and Pierre-Louis Lions (1992), User's guide to viscosity solutions of second-order partial differential equations, {\it Bulletin of the American Mathematical Society}, 27 (1): 1-67.

\smallskip \noindent \hangindent 20 pt Dubins, Lester E. and Leonard J. Savage (1965, 1976), {\it How to Gamble if You Must: Inequalities for Stochastic Processes}, 1965 edition McGraw-Hill, New York. 1976 edition Dover, New York.

%\smallskip \noindent \hangindent 20 pt Fasano, Antonio and Mario Primicerio (1997), General free-boundary problems for the heat equation, III, {\it Journal of Mathematical Analysis and Applications}, 59: 1-14.

\smallskip \noindent \hangindent 20 pt Karatzas, Ioannis (1997), Adaptive control of a diffusion to a goal, and a parabolic Monge-Ampere-type equation, {\it Asian Journal of Mathematics}, 1: 295-313.

\smallskip \noindent \hangindent 20 pt Karatzas, Ioannis and Steven E. Shreve (1998), {\it Methods of Mathematical Finance}, New York: Springer-Verlag.

\smallskip \noindent \hangindent 20 pt Kulldorff, Martin (1993), Optimal control of favorable games with a time limit, {\it SIAM Journal on Control and Optimization}, 31 (1): 52-69.

%\smallskip \noindent \hangindent 20 pt Merton, Robert C. (1969), Lifetime portfolio selection under uncertainty: the continuous-time case, {\it Review of Economics and Statistics}, 51 (3): 247-257.

%\smallskip \noindent \hangindent 20 pt Milevsky, Moshe A., Kristen S. Moore, and Virginia R. Young (2006), Asset allocation and annuity-purchase strategies to minimize the probability of financial ruin, {\it Mathematical Finance}, 16 (4): 647-671. 

%\smallskip \noindent \hangindent 20 pt Moore, Kristen S. and Virginia R. Young (2006), Optimal and simple, nearly-optimal rules for minimizing the probability of financial ruin in retirement, {\it North American Actuarial Journal}, 10 (4): 145Ð161.

\smallskip \noindent \hangindent 20 pt Orey, Steven, Victor C. Pestien, and William D. Sudderth (1987), Reaching zero rapidly, {\it SIAM Journal on Control and Optimization} 25 (5): 1253-1265.

\smallskip \noindent \hangindent 20 pt Pestien, Victor C. and William D. Sudderth (1985), Continuous-time red and black: how to control a diffusion to a goal, {\it Mathematics of Operations Research}, 10 (4): 599-611.

\smallskip \noindent \hangindent 20 pt Pliska, Stanley and Jinchun Ye (2007), Optimal life insurance purchase and consumption/investment under uncertain lifetime, {\it Journal of Banking and Finance}, 31 (5): 1307-1319.

\smallskip \noindent \hangindent 20 pt Promislow, S. David and Virginia R. Young (2005), Minimizing the probability of ruin when claims follow Brownian motion with drift, {\it North American Actuarial Journal}, 9 (3): 109-128.

%\smallskip \noindent \hangindent 20 pt Polyanin, Andrei D. and Valentin F. Zaitsev (2003), {\it Handbook of Exact Solutions for Ordinary Differential Equations}, second edition, Chapman and Hall/CRC, Boca Raton.

\smallskip \noindent \hangindent 20 pt Richard, Scott F. (1975), Optimal consumption, portfolio and life insurance rules for an uncertain lived individual in a continuous time model, {\it Journal of Financial Economics}, 2 (2): 187-203.

\smallskip \noindent \hangindent 20 pt Schmidli, Hanspeter (2002), On minimizing the ruin probability of investment and reinsurance, {\it Journal of Applied Probability}, 12 (3): 890-907.

\smallskip \noindent \hangindent 20 pt Sethi, Suresh P., Michael I. Taksar, and Ernst L. Presman (1992), Explicit solution of a general consumption/portfolio problem with subsistence consumption and bankruptcy, {\it Journal of Economic Dynamics and Control}, 16 (3-4): 747-768.

\smallskip \noindent \hangindent 20 pt Sudderth, William D. and Ananda Weerasinghe (1989), Controlling a process to a goal in finite time, {\it Mathematics of Operations Research}, 14 (3): 400-409.

\smallskip \noindent \hangindent 20 pt Vellekoop, Michel H. and Mark H. A. Davis (2009), An optimal investment problem with randomly terminating income, Joint 48th IEEE Conference on Decision and Control, 3650-3655. Also available via the Amsterdam School of Economics Research Institute at \hfill \break http://dare.uva.nl/document/2/95537.

%\smallskip \noindent \hangindent 20 pt Walter, Wolfgang (1970), {\it Differential and Integral Inequalities}, New York: Springer-Verlag.

\smallskip \noindent \hangindent 20 pt Wang, Ting and Virginia R. Young (2012), Optimal commutable annuities to minimize the probability of lifetime ruin, {\it Insurance: Mathematics and Economics}, 50 (1): 200-216.

%\smallskip \noindent \hangindent 20 pt Wang, Ting and Virginia R. Young (2012b), Maximizing the utility of consumption with commutable annuities, {\it Insurance: Mathematics and Economics}, 51 (2): 352-369.

\smallskip \noindent \hangindent 20 pt Young, Virginia R. (2004), Optimal investment strategy to minimize the probability of lifetime ruin, {\it North American Actuarial Journal}, 8 (4): 105-126.

 \bye